\documentclass[secnumarabi,superscriptaddress,amssymb,nobibnotes,aps]{revtex4}

\usepackage{epsfig}
\usepackage{graphicx}
\usepackage{amsmath}
\usepackage{hyperref}

\usepackage{color}

\newcommand \Pomeron {I\!\!P}

\begin{document} 

\title{Suppression of diffraction in deep-inelastic scattering on nuclei and dynamical mechanism of leading twist nuclear shadowing}

\author{V. Guzey}

\affiliation{University  of  Jyvaskyla, Department of Physics,  P.O. Box 35, FI-40014  University  of  Jyvaskyla,  Finland} 
\affiliation{Helsinki Institute of Physics, P.O.  Box  64,  FI-00014  University  of  Helsinki,  Finland} 

\author{M. Strikman}
\affiliation{Department of Physics, The Pennsylvania State University, State College, PA 16802, USA}

\pacs{} 

\begin{abstract} 

Using the leading twist approach (LTA) to nuclear shadowing, we calculate the ratios of diffractive and usual parton distributions for a heavy nucleus (Pb) and the proton, 
$R_{A/p}=(f_{i/A}^{D(3)}/f_{i/A})/(f_{i/p}^{D(3)}/f_{i/p})$, for coherent and summed (coherent plus quasi-elastic) nuclear deep-inelastic scattering. We find
that $R_{A/p} \approx 0.5-1$ for quarks as well as for the ratio of the diffractive and total cross sections $[(d\sigma_{\rm diff}/dM_X^2)/\sigma_{\rm tot}]_{eA}/[(d\sigma_{\rm diff}/dM_X^2)/\sigma_{\rm tot}]_{ep}$ and $R_{A/p} \approx 0.5-1.3$ for gluons 
in a broad range of $x$, including the kinematics of the Electron-Ion Collider,
which reaffirms the difference from the nuclear enhancement of $R_{A/p}$ predicted in the gluon saturation framework. 
We demonstrate that the magnitude of $R_{A/p}$ is controlled by the cross section of the interaction of hadronic fluctuations of the virtual photon with target nucleons, which explains an enhancement of $R_{A/p}$ in the color dipole model and its suppression in LTA. We argue that the black disk limit 
corresponds to $R_{A/p}=1$ and $R^{\rm coh}_{A/p}=0.86$ for the summed and coherent scattering, respectively. Relying on an intuitive definition of the 
saturation scale, we show that the ratio of the saturation scales of a heavy nucleus and proton $Q_{sA}^2(b)/Q_{sp}^2(b) \approx 1$ at small impact parameters $b$
due to the strong leading twist nuclear shadowing and diluteness of the nuclear density.

\end{abstract}

\maketitle 

\section{Introduction}
\label{sec:intro}

One of the overarching goals of high energy nuclear physics is to understand the microscopic structure of nuclei and nucleons 
and the dynamics of strong interactions in terms of quantum chromodynamics (QCD). On the one hand, a wide array of hard scattering
processes involving nuclei can be interpreted in terms of cold nuclear matter effects, which imply nuclear modifications of
quark and gluon (parton) distribution functions (PDFs), for a recent review, see Ref.~\cite{Klasen:2023uqj}. 
On the other hand, one continues to search for new states of matter in QCD at increasingly high energies, which are characterized by very high parton densities leading to their non-linear dynamics and saturation and in general a different effective description~\cite{Gelis:2010nm,Morreale:2021pnn}.

Within the framework of collinear factorization and perturbative QCD~\cite{CTEQ:1993hwr}, global analyses of data on lepton-nucleus deep inelastic scattering (DIS) in fixed-target experiments and proton-nucleus scattering at the Relativistic Heavy Ion Collider (RHIC) and the Large Hadron Collider (LHC) 
 have shown that the suppression of the nuclear cross sections (structure functions) is translated into the nuclear suppression
 (shadowing) of quark and gluon PDFs for small $x < 0.05$, where $x$ is the parton momentum fraction~\cite{Eskola:2021nhw,AbdulKhalek:2022fyi,Kovarik:2015cma}. Alternatively, the nuclear shadowing effect for quark and gluon PDFs can be calculated using the leading twist approach (LTA), which combines information on the low-energy nuclear structure, methods of soft hadron-nucleus scattering, and QCD factorization theorems for inclusive and diffractive DIS~\cite{Frankfurt:2011cs}.
In both cases nuclear shadowing is predicted to be a leading 
 twist effect, which has a weak, logarithmic dependence on the photon virtuality $Q^2$ (hard scale of a process).
 
At the same time, the data on fixed-target nuclear DIS can also be described assuming that nuclear shadowing is an effect of power-suppressed, higher-twist corrections~\cite{Qiu:2003vd} or a nuclear enhancement of the saturation scale in the color glass condensate (CGC) framework~\cite{Kowalski:2007rw}, for a review, see~\cite{Armesto:2006ph}. It raises the question of the dynamical mechanism of 
nuclear shadowing and its distinction from saturation. Note that while the leading twist picture implies sufficiently large $Q^2$ and finite $x$ (Bjorken limit) and the saturation framework is developed in the limit of very small $x$ and finite $Q^2$ (Regge limit),
the above question is valid in the common region of applicability of the two approaches, which overlaps with the kinematic coverage of the past, present and near-future measurements.

It was argued in~\cite{Frankfurt:2002kd} that to discriminate between the leading-twist and higher-twist descriptions of nuclear shadowing,
it is advantageous to study the $Q^2$ dependence of observables dominated by small-size partonic fluctuations (color dipoles) of the virtual photon, notably, the nuclear longitudinal structure function $F_L^A(x,Q^2)$. Indeed, in the leading twist picture, the high-energy probe resolves the nuclear partonic structure by simultaneously coupling to target nucleons through diffractive exchanges, which allows one to express nuclear shadowing for nuclear PDFs of a given flavor (quark or gluon) in terms of the nucleon leading-twist diffractive PDF of the same flavor.
 In the infinite momentum frame, it corresponds to interference of diffractive scattering off two different nucleons in the $|in \rangle$ and $\langle out |$ states~\cite{Frankfurt:1998ym}.
At the same time, the saturation mechanism of nuclear shadowing is usually realized through successive interactions of quark-antiquark (and sometimes
also quark-antiquark-gluon) dipoles with target nucleons leading to a reduction of the nuclear cross section. 
Selecting a special observable, e.g., the longitudinal DIS structure function or the cross section of heavy quarkonium production, 
when the dipole cross section is dominated by small-size dipoles and hence scales as $1/Q^2$, one finds that nuclear shadowing also scales as $1/Q^2$, i.e., 
it gives a higher-twist correction.

An argument spectacularly supporting the leading twist mechanism of nuclear shadowing was given in Refs.~\cite{Guzey:2013xba,Guzey:2013qza}, which showed that
coherent $J/\psi$ photoproduction in Pb-Pb ultraperipheral collisions (UPCs) at the LHC gives direct evidence of strong gluon nuclear shadowing at $x \approx 10^{-3}$ predicted
by LTA~\cite{Frankfurt:2011cs}. Recent measurements of $J/\psi$ photoproduction in heavy-ion UPCs at 
the LHC~\cite{ALICE:2021gpt,ALICE:2019tqa,LHCb:2022ahs,CMS:2023snh,ALICE:2023jgu} and RHIC~\cite{STAR:2023gpk} 
further confirmed these predictions and extended the kinematic coverage down to $x \approx 10^{-5}$. Note, however, that the interpretation of these data
in terms of gluon nuclear shadowing is complicated by very large next-to-leading order (NLO) corrections~\cite{Eskola:2022vpi,Eskola:2022vaf}.
Alternatively, these UPC data at small $x$ have been described in a specific realization of the dipole dipole including the non-linear saturation effects in the dipole
cross section~\cite{Bendova:2020hbb,Cepila:2017nef}. It can be taken as a sign that coherent $J/\psi$ photoproduction on nuclei might not have a sufficient
discriminating power to distinguish among different mechanisms of nuclear shadowing because of significant power-suppressed and relativistic corrections
to the charmonium light-cone wave function~\cite{Frankfurt:1997fj,Lappi:2020ufv}.

In addition, it has recently been suggested in the literature that novel signals of saturation in UPCs can be searched for in semi-inclusive photoproduction jets in diffractive nucleus-nucleus scattering~\cite{Iancu:2023lel}  and in cross section ratios of elastic vector meson photoproduction to inclusive hadron or jet photoproduction in heavy-ion and proton-nucleus scattering~\cite{Kovchegov:2023bvy}.
Note that one of the challenges of the proposed jet measurement is the need to detect jets with transverse momenta of a few GeV,
otherwise the LHC kinematics, detector acceptance, and luminosity do not allow to reach $x \lesssim 10^{-3}$, see also~\cite{Guzey:2016tek,Guzey:2018dlm}.

In the context of the planned Electron-Ion Collider in the U.S.~\cite{Accardi:2012qut}, it has been emphasized that a  process sensitive to small-$x$
QCD dynamics is inclusive diffraction in lepton-nucleus DIS. In particular, the ratio of the diffractive to the total cross sections for a heavy nucleus and the proton, $R_{A/p}=[(d\sigma_{\rm diff}/dM_X^2)/\sigma_{\rm tot}]_{eA}/[(d\sigma_{\rm diff}/dM_X^2)/\sigma_{\rm tot}]_{ep}$, where $M_X$ is the mass of the diffractively produced final state, has been put forward as a promising observable: the ratio $R_{A/p} \approx 1.5-2$ in the saturation framework~\cite{Kowalski:2007rw,Lappi:2023frf} and $R_{A/p} < 1$ in LTA~\cite{Frankfurt:2011cs}.

The aim of the present paper is to revisit and update the LTA predictions for $R_{A/p}$, highlight their interpretation and comparison to
the competing results of the saturation framework. We reiterate the LTA observation that strong leading twist nuclear shadowing significantly
suppresses the quark and gluon nuclear diffractive PDFs compared to their impulse approximation (IA) estimates.
Combining it with the LTA predictions for usual nuclear PDFs, we find that $R_{A/p} \approx 0.5-1$ for quarks and for $[(d\sigma_{\rm diff}/dM_X^2)/\sigma_{\rm tot}]_{eA}/[(d\sigma_{\rm diff}/dM_X^2)/\sigma_{\rm tot}]_{ep}$
and $R_{A/p} \approx 0.5-1.3$ for gluons.
These results depend weakly on $x$, do not depend on $M_X$ (the Pomeron momentum fraction $x_{\Pomeron}$), and are characterized by 
a significant theoretical uncertainty due to modeling of the effective rescattering cross section $\sigma_{\rm soft}$.
We demonstrate that the magnitude of $R_{A/p}$ critically depends on $\sigma_{\rm soft}$: taking $\sigma_{\rm soft} \sim \sigma_{\rho N}=20-30$ mb, where 
$\sigma_{\rho N}$ is the $\rho$ meson-nucleon cross section, which is typical for the dipole model,
leads to an enhanced $R_{A/p} \approx 1.5-2$, while larger $\sigma_{\rm soft} \approx
40-50$ mb, which are characteristic for the full-fledged leading twist shadowing, corresponds to $R_{A/p} \approx 0.5-1.3$. The spread of the predicted values depends on 
$\sigma_{\rm soft}$ and its uncertainty, which can also be interpreted in terms of point-like fluctuations of the virtual photon that are not suppressed by shadowing.
Thus, we show that the magnitude of $R_{A/p}$ is sensitive to the underlying strength of the interaction of hadronic fluctuations (color dipoles)
of the virtual photon with target nucleons and not so much to the gluon saturation. In the black disk limit for these interactions, which 
satisfies $S$-channel unitarity for the dipole-nucleon cross section,
$R_{A/p}^{\rm coh}=0.86$ in the case of purely coherent scattering and $R_{A/p}=1$ for a sum of coherent and quasi-elastic (total rapidity gap) contributions.
We argue that the leading twist nuclear shadowing slows down an onset of gluon saturation and illustrate it by showing 
that the ratio of the saturation scales of a heavy nucleus and proton $Q_{sA}^2(b)/Q_{sp}^2(b) \approx 1$ at small impact parameters $b$.
Part of this absence of a nuclear enhancement of $Q_{sA}^2(b)/Q_{sp}^2(b)$ is also caused by the relative diluteness of the nuclear gluon distribution in the transverse plane compared to that in the proton, which is driven by the realistic  nuclear density.

The rest of this paper is organized as follows. In Sec.~\ref{sec:lta}, we recapitulate the derivation of the nuclear diffractive 
structure functions and PDFs within the leading twist approach and make numerical predictions for the ratios of the nucleus and proton quark 
and gluon diffractive PDFs, $f_{i/A}^{D(3)}/[Af^{D(3)}_{i/p}]$, for the coherent and summed (coherent plus quasi-elastic) nuclear DIS.
We formulate and discuss the LTA predictions for the ratios of the diffractive and usual PDFs for a heavy nucleus and the proton,
$R^{\rm coh}_{A/p}$ and $R_{A/p}$, including the case of the nucleon black disk limit, in Sec.~\ref{sec:super-ratio}. In Sec.~\ref{sec:saturation}, we examine the 
influence of the leading twist nuclear shadowing on the absence (significant reduction) of a nuclear enhancement of the saturation scale.
We summarize our results in Sec.~\ref{sec:conclusions}.

\section{Leading twist approximation for nuclear diffractive PDFs}
\label{sec:lta}

\begin{figure}[t]
\centerline{%
\includegraphics[width=17cm]{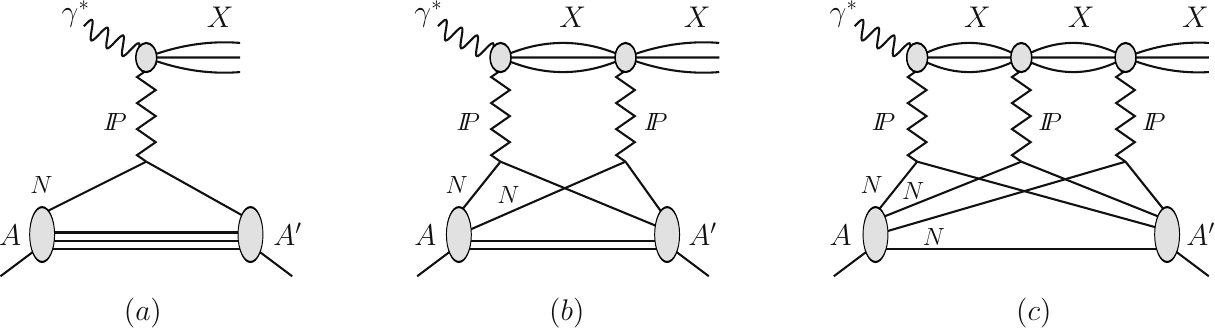}
}
\caption{The $\gamma^{\ast}+A \to X+A^{\prime}$ amplitude of Eq.~(\ref{eq:diff1}) as a series of interactions with one (graph $a$), two (graph $b$), and 
three (graph $c$) nucleons of a nuclear target $A$. The zigzag lines labeled ``$\Pomeron$'' denote diffractive (Pomeron) exchanges producing the state $X$.
The final nuclear state $A^{\prime}$ includes both coherent ($A^{\prime}=A$) and quasi-elastic ($A^{\prime} \neq A$) scattering.}
\label{fig:Diff_2024}
\end{figure}

The leading twist approximation (LTA) to nuclear shadowing combines methods of high energy hadron-nucleus scattering with QCD collinear factorization theorems
for inclusive~\cite{CTEQ:1993hwr} and diffractive~\cite{Collins:1997sr} DIS. The $\gamma^{\ast}+A \to X+A^{\prime}$ amplitude for inclusive diffractive DIS on a nucleus can be presented as a series, where each term corresponds to the interaction with $i=1,2, \dots, A$ nucleons, see Fig.~\ref{fig:Diff_2024},
\begin{equation}
\Gamma_{\gamma^{\ast}A \to XA^{\prime}}(\vec{b})=\langle A^{\prime}|\sum_{i}^{A} \Gamma_{\gamma^{\ast}N \to XN}(\vec{b}-{\vec r}_{i \perp}) e^{i z_i \Delta_{\gamma^{\ast}X}}
\prod_{j \neq i} \left(1-\theta(z_j-z_i) \Gamma_X(\vec{b}-{\vec r}_{j \perp}) \right)|A \rangle \,.
\label{eq:diff1}
\end{equation}
In Eq.~(\ref{eq:diff1}),  $\Gamma_{\gamma^{\ast}N \to XN}$ is the amplitude of diffractive scattering on nucleon $i$ and $\Gamma_{X}$ is the amplitude of soft scattering
of the diffractive state $X$ on remaining nucleons $j \neq i$; the sum of the latter contributions produces the nuclear shadowing effect. All amplitudes are 
written in the coordinate representation, where $\vec{b}$ is the impact parameter, ${\vec r}_{i \perp}$ and $z_i$ are the transverse and longitudinal positions
of the nucleons. The Heaviside step function $ \theta(z_j-z_i)$ takes into account the space-time development of the process in the 
nucleus rest frame
and $\Delta_{\gamma^{\ast}X}=x_{\Pomeron} m_N$ is the non-zero longitudinal momentum transfer in the $\gamma^{\ast}+ N \to X+N$ process on the nucleon, where $x_{\Pomeron}$ is a small momentum fraction carried by the diffractive exchange (Pomeron)
and $m_N$ is the nucleon mass. The initial and final nuclear states are denoted by $A$ and $A^{\prime}$, respectively.
Note that we assume that the states $X$ stay ``frozen'' during their passage through the nucleus and, thus, neglect possible $X \to X^{\prime}$,
$X^{\prime} \to X^{\prime \prime}$, etc.~non-diagonal transitions.

The scattering amplitudes in Eq.~(\ref{eq:diff1}) are taken in the standard form tailored to hadron-nucleus scattering at high energies~\cite{Bauer:1977iq}.
For the diffractive amplitude, we use
\begin{equation}
\Gamma_{\gamma^{\ast}N \to XN}(\vec{b})=\frac{1-i \eta}{4 \pi B_{\rm diff}\sqrt{1+\eta^2}} \sqrt{16 \pi \frac{d\sigma_{\gamma^{\ast} N \to XN}(t=0)}{dt}} e^{-b^2/(2 B_{\rm diff})} \,,
\label{eq:amp1}
\end{equation}
where $B_{\rm diff}=6$ GeV$^{-2}$ is the slope of the $t$-dependence of the diffractive DIS with a leading proton at HERA~\cite{H1:2006uea},
and $\eta \approx \pi/2 (\alpha_{\Pomeron}(0)-1)=0.17$ is the ratio of the real and imaginary part of this amplitude, which is estimated using 
the dispersion relation and the Pomeron intercept $\alpha_{\Pomeron}(0)=1.111$~\cite{H1:2006zyl}. With these conventions, 
$\sigma_{\gamma^{\ast} N \to XN}=\int d^2 \vec{b}\, |\Gamma_{\gamma^{\ast}N \to XN}(\vec{b})|^2= (d\sigma_{\gamma^{\ast} N \to XN}(t=0)/dt)/B_{\rm diff}$.

The amplitude for the soft scattering is 
\begin{equation}
\Gamma_{X}(\vec{b})=(1-i \eta)\frac{\sigma_{\rm soft}}{4 \pi B_{\rm X}} e^{-b^2/(2 B_X)} \,,
\label{eq:amp2}
\end{equation}
where $\sigma_{\rm soft}$ and $B_X$ are the corresponding total cross section and the slope of the $t$-dependence.  Note that while all parameters in
Eq.~(\ref{eq:amp1}) are constrained by HERA measurements of diffraction in lepton-proton DIS, the soft cross section $\sigma_{\rm soft}$ and to a lesser degree
$B_X$ can only be estimated and need to be modeled, see the discussion in Sec.~\ref{subsec:results}. For simplicity of the resulting expressions and their transparent physical interpretation, we take
$B_X=B_{\rm diff}$ below.

\subsection{Coherent scattering}
\label{subsec:coherent}

In the case of coherent nuclear diffraction, $A =A^{\prime}$. Using the model of independent nucleons for the nuclear wave function, which is commonly used for heavy nuclei, one obtains from 
Eq.~(\ref{eq:diff1})
\begin{equation}
\Gamma_{\gamma^{\ast}A \to XA}(\vec{b})=\frac{1-i \eta}{2\sqrt{1+\eta^2}} \sqrt{16 \pi \frac{d\sigma_{\gamma^{\ast} N \to XN}(t=0)}{dt}} 
\int dz \rho_A(\vec{b},z)
 e^{i z \Delta_{\gamma^{\ast}X}} e^{-\frac{1-i \eta}{2} \sigma_{\rm soft}\int_z^{\infty} dz^{\prime} \rho_A(\vec{b},z^{\prime})} \,,
\label{eq:diff2}
\end{equation}
where $\rho_A(\vec{r})$ is the nuclear density~\cite{DeVries:1987atn} normalized to the nucleus mass number, $\int d^3 \vec{r} \rho_A(\vec{r})=A$. The derivation of Eq.~(\ref{eq:diff2}) uses that $B_{\rm diff} \ll R_A^2$ and  $B_X \ll R_A^2$, where $R_A$ is the nucleus radius, which places all the nucleons at the same impact parameter $\vec{b}$.
The effect of the interaction with $j \neq i$ nucleons is eikonalized and written in an exponential form, which is a good approximation for large $A$.

Using Eq.~(\ref{eq:diff2}), one obtains for the $t$-integrated cross section of coherent diffraction in $\gamma^{\ast}A$ scattering,
\begin{equation}
\sigma_{\gamma^{\ast}A \to XA}=\int d^2 \vec{b}\, |\Gamma_{\gamma^{\ast}A \to XA}(\vec{b})|^2=4 \pi \frac{d\sigma_{\gamma^{\ast} N \to XN}(t=0)}{dt}
\int d^2 \vec{b}\,\left| \int dz \rho_A(\vec{b},z)
 e^{i z \Delta_{\gamma^{\ast}X}} e^{-\frac{1-i \eta}{2} \sigma_{\rm soft}\int_z^{\infty} dz^{\prime} \rho_A(\vec{b},z^{\prime})} \right|^2 \,.
\label{eq:diff3}
\end{equation}
Note that this expression is rather general and valid in both soft (photoproduction and low $Q^2$) and hard (high $Q^2$) regimes
of $\gamma^{\ast}A$ scattering. In the latter case, one finds for the nuclear diffractive structure function $F_{2A}^{D(3)}$,
\begin{equation}
F_{2A}^{D(3)}(x,x_{\Pomeron},Q^2)=4 \pi B_{\rm diff} F_{2p}^{D(3)}(x,x_{\Pomeron},Q^2)
\int d^2 \vec{b}\,\left| \int dz \rho_A(\vec{b},z)
 e^{i z x_{\Pomeron} m_N} e^{-\frac{1-i \eta}{2} \sigma_{\rm soft}(x)\int_z^{\infty} dz^{\prime} \rho_A(\vec{b},z^{\prime})} \right|^2 \,,
\label{eq:diff4}
\end{equation}
where $F_{2p}^{D(3)}$ is the free nucleon (proton) diffractive structure function, which has been extensively measured at HERA. In Eq.~(\ref{eq:diff4}),
we explicitly show the dependence of the involved quantities on $x$, $x_{\Pomeron}$ and $Q^2$. Note that we neglected the possible dependence
of $\sigma_{\rm soft}$ on $x_{\Pomeron}$. This is motivated by the observation that the fraction of diffractive events in the total DIS cross section is
a weak function of the diffractive mass $M_X$~\cite{H1:2006zyl}. 
We also suppressed the explicit dependence of $\sigma_{\rm soft}$ on $Q^2$ because LTA is meant for the calculation of nuclear structure functions and PDFs at $Q^2={\cal O}(\rm few)$ GeV$^2$, which serve as an initial condition (input) for the subsequent Dokshitzer-Gribov-Lipatov-Altarelli-Parisi (DGLAP) $Q^2$ evolution to higher resolution scales. Our numerical analysis below implicitly assumes that $Q^2=Q_0^2=4$ GeV$^2$.
This value is a compromise between the needs to have $Q_0$ sufficiently low for the applicability of methods of cross section fluctuations
for soft hadron-nucleus scattering~\cite{Guzey:2009jr} and sufficiently high to minimize the influence of higher-twist effects in
diffraction in lepton-proton DIS~\cite{Golec-Biernat:2007mao,Salajegheh:2022vyv}. Note that the choice of lower values of $Q_0$, e.g.,
$Q_0^2=2.5$ GeV$^2$, is also possible and leads to consistent predictions for nuclear PDFs at $Q^2=4$ GeV$^2$, see Sec.~5.9 of~\cite{Frankfurt:2011cs}.

Considering the limit of small $x_{\Pomeron}$ (not very large diffractive masses $M_X$), one can neglect the longitudinal momentum transfer (the  $e^{i z x_{\Pomeron} m_N}$ factor) in Eq.~(\ref{eq:diff4}) and obtain after integration by parts
\begin{eqnarray}
F_{2A}^{D(3)}(x,x_{\Pomeron},Q^2) &=& 16 \pi B_{\rm diff} F_{2p}^{D(3)}(x,x_{\Pomeron},Q^2)
\int d^2 \vec{b}\,\left| \frac{1-e^{-\frac{1-i \eta}{2} \sigma_{\rm soft}(x)T_A(\vec{b})}}{(1-i\eta)\sigma_{\rm soft}(x)} \right|^2 \nonumber\\
&=&  F_{2p}^{D(3)}(x,x_{\Pomeron},Q^2) \frac{1}{\sigma_{\rm el}(x)} \int d^2 \vec{b}\,\left|1-e^{-\frac{1-i \eta}{2} \sigma_{\rm soft}(x)T_A(\vec{b})}\right|^2 
\,,
\label{eq:diff5}
\end{eqnarray}
where $T_A(\vec{b})=\int dz \rho(\vec{b},z)$ is the so-called nuclear optical density and $\sigma_{\rm el}(x)$ is the elastic cross section
\begin{equation}
\sigma_{\rm el}(x)=(1+\eta^2) \frac{[\sigma_{\rm soft}(x)]^2}{16 \pi B_{\rm diff}} \,.
\label{eq:sigma_el}
\end{equation}
This form of Eq.~(\ref{eq:diff5}) allows for its straightforward interpretation: the nuclear shadowing effect for $F_{2A}^{D(3)}/F_{2p}^{D(3)}$
is given by the ratio of the nuclear and nucleon elastic cross sections of the hadronic fluctuations of the virtual photon, which
compose the diffractive final state $X$.

The main feature of LTA is the use of the QCD factorization theorem for diffractive hard scattering~\cite{Collins:1997sr}, 
which allows one to convert predictions for nuclear cross sections and structure functions into those for individual PDFs. Applying the QCD factorization theorem to
Eq.~(\ref{eq:diff5}), one obtains for the nuclear diffractive PDFs $f_{i/A}^{D(3)}$,
\begin{equation}
f_{i/A}^{D(3)}(x,x_{\Pomeron},Q^2) =  f_{i/p}^{D(3)}(x,x_{\Pomeron},Q^2) \frac{1}{\sigma_{\rm el}^i(x)} \int d^2 \vec{b}\,\left|1-e^{-\frac{1-i \eta}{2} \sigma_{\rm soft}^i(x)T_A(\vec{b})}\right|^2 \,,
\label{eq:diff6}
\end{equation}
where $f_{i/p}^{D(3)}$ are the proton diffractive PDFs and $i$ is the parton flavor (quark or gluon). Note that we introduced the explicit dependence of $\sigma_{\rm soft}$ and $\sigma_{\rm el}$ 
on the parton flavor because the interaction strength (the proton diffractive PDFs) are very different in the quark and gluon channels~\cite{H1:2006zyl}. 
For a more general expression of $f_{i/A}^{D(3)}$ with $e^{i z x_{\Pomeron} m_N} \neq 1$, see Ref.~\cite{Frankfurt:2003gx}.

\subsection{Quasi-elastic scattering}
\label{subsec:quasi-elastic}

Besides purely elastic scattering $A=A^{\prime}$, rapidity gap events in $\gamma^{\ast}A$ scattering also include nuclear quasi-elastic scattering corresponding to $A^{\prime} \neq A$. Using completeness of nuclear final states, the sum of the elastic and quasi-elastic cross sections can be written as
\begin{equation}
\sigma_{\gamma^{\ast}A \to XA^{\prime}}=\int d^2 \vec{b} \sum_{A^{\prime}} \langle A| \Gamma_{\gamma^{\ast}A \to XA}^{\dagger}(\vec{b})|A^{\prime} \rangle
\langle A^{\prime}| \Gamma_{\gamma^{\ast}A \to XA}(\vec{b})|A \rangle = \int d^2\vec{b}\, \langle A| \left|\Gamma_{\gamma^{\ast}A \to XA}(\vec{b})\right|^2|A \rangle \,,
\label{eq:inc1}
\end{equation}
where $\Gamma_{\gamma^{\ast}A \to XA}(\vec{b})$ is given by Eq.~(\ref{eq:diff1}). When squaring the scattering amplitude in Eq.~(\ref{eq:inc1}), one encounters two types of terms: diffractive scattering can take place on different nucleons (similarly to the purely coherent case) or on one nucleon (incoherent scattering),
\begin{eqnarray}
\langle A| \left|\Gamma_{\gamma^{\ast}A \to XA}(\vec{b})\right|^2|A \rangle &=& \langle A|\sum_{i \neq j}^{A} \Gamma^{\dagger}_{\gamma^{\ast}N \to XN}(\vec{b}-{\vec r}_{j \perp}) \Gamma_{\gamma^{\ast}N \to XN}(\vec{b}-{\vec r}_{i \perp}) e^{i (z_i-z_j) \Delta_{\gamma^{\ast}X}} \nonumber\\
& \times & \prod_{j^{\prime} \neq j} \left(1-\theta(z_{j^{\prime}}-z_j) \Gamma_X^{\dagger}(\vec{b}-{\vec r}_{j^{\prime} \perp})\right)
\prod_{i^{\prime} \neq i } \left(1-\theta(z_{i^{\prime}}-z_i) \Gamma_X(\vec{b}-{\vec r}_{i^{\prime} \perp})\right)|A \rangle \nonumber\\
&+&\langle A|\sum_{i}^{A} |\Gamma_{\gamma^{\ast}N \to XN}(\vec{b}-{\vec r}_{i \perp})|^2 \prod_{i^{\prime} \neq i} \left|1-\theta(z_{i^{\prime}}-z_i) \Gamma_X(\vec{b}-{\vec r}_{i^{\prime} \perp})\right|^2|A \rangle \,.
\label{eq:inc1_b}
\end{eqnarray}
 For the first term in Eq.~(\ref{eq:inc1_b}), one obtains after some algebra
\begin{eqnarray}
&& \langle A| \left|\Gamma_{\gamma^{\ast}A \to XA}(\vec{b})\right|^2|A \rangle_{\rm coh} = 4 \pi \frac{d\sigma_{\gamma^{\ast} N \to XN}(t=0)}{dt}
 \int^{\infty}_{-\infty} dz_1 \rho_A(\vec{b},z_1) \int_{z_1}^{\infty} dz_2 \rho_A(\vec{b},z_2)  \nonumber\\
 &\times& \left[e^{i (z_1-z_2) \Delta_{\gamma^{\ast}X}}\left(1-\frac{(1-i \eta)\sigma_{\rm soft}}{8 \pi B_{\rm diff}}\right)
  e^{-\frac{1-i \eta}{2} \sigma_{\rm soft}\int_{z_1}^{\infty} dz^{\prime} \rho_A(\vec{b},z^{\prime})-(\frac{1+i \eta}{2} \sigma_{\rm soft}-\sigma_{\rm el})\int_{z_2}^{\infty} dz^{\prime} \rho_A(\vec{b},z^{\prime})}+{\rm C.T.} \right] \,.
 \label{eq:inc2}
\end{eqnarray}
Taking the $\Delta_{\gamma^{\ast}X} \to 0$ limit, one finds after integration by parts
\begin{eqnarray}
 \langle A| \left|\Gamma_{\gamma^{\ast}A \to XA}(\vec{b})\right|^2|A \rangle_{\rm coh} &=& 16 \pi \frac{d\sigma_{\gamma^{\ast} N \to XN}(t=0)}{dt} \frac{1}{(1+\eta^2)\sigma_{\rm soft}^2} \nonumber\\
 &\times& \left[1-2 \Re e\, e^{-\frac{1-i \eta}{2} \sigma_{\rm soft} T_A(\vec{b})}+e^{-\sigma_{\rm in} T_A(\vec{b})}-\frac{\sigma_{\rm el}}{\sigma_{\rm in}}
 \left(1-e^{-\sigma_{\rm in} T_A(\vec{b})} \right) \right] \nonumber\\
 &=&\sigma_{\gamma^{\ast} N \to XN} \frac{1}{\sigma_{\rm el}} \left[1-2 \Re e\, e^{-\frac{1-i \eta}{2} \sigma_{\rm soft} T_A(\vec{b})}+e^{-\sigma_{\rm in} T_A(\vec{b})}-\frac{\sigma_{\rm el}}{\sigma_{\rm in}}
 \left(1-e^{-\sigma_{\rm in} T_A(\vec{b})} \right) \right] 
 \,,
 \label{eq:inc2_b}
\end{eqnarray}
where $\sigma_{\rm in}=\sigma_{\rm soft}-\sigma_{\rm el}$.

For the second contribution in Eq.~(\ref{eq:inc1_b}) [the last line in Eq.~(\ref{eq:inc1_b})], one obtains
\begin{eqnarray}
 \langle A| \left|\Gamma_{\gamma^{\ast}A \to XA}(\vec{b})\right|^2|A \rangle_{\rm incoh} &=& \frac{d\sigma_{\gamma^{\ast} N \to XN}(t=0)}{dt}\frac{1}{B_{\rm diff}}\int dz \rho_A(\vec{b},z) e^{-\sigma_{\rm in} \int_z^{\infty} dz^{\prime} \rho_A(\vec{b},z^{\prime})} \nonumber\\
 &=& \sigma_{\gamma^{\ast} N \to XN} \frac{1}{\sigma_{\rm in}}\left(1- e^{-\sigma_{\rm in}T_A(\vec{b})}\right) \,.
\label{eq:inc3}
\end{eqnarray}
Combining Eqs.~(\ref{eq:inc2_b}) and (\ref{eq:inc3}), we obtain for the $t$-integrated total cross section of diffraction in $\gamma^{\ast}A$ scattering
\begin{equation}
\sigma_{\gamma^{\ast}A \to XA^{\prime}}=\sigma_{\gamma^{\ast} N \to XN} \frac{1}{\sigma_{\rm el}}\int d^2 \vec{b} \left(\left|1-e^{-\frac{1-i \eta}{2} \sigma_{\rm soft} T_A(\vec{b})}\right|^2+e^{-\sigma_{\rm in} T_A(\vec{b})}-e^{-\sigma_{\rm soft} T_A(\vec{b})} \right) \,. 
\label{eq:inc4} 
\end{equation}
In this equation, the first term corresponds to coherent scattering, see Eq.~(\ref{eq:diff5}), and the second term is the contribution of quasi-elastic (incoherent)  nuclear scattering. In the regime of hard scattering,
one can introduce the corresponding nuclear diffractive structure function
\begin{equation}
\tilde{F}^{D(3)}_{2A}(x,x_{\Pomeron},Q^2)=F^{D(3)}_{2p}(x,x_{\Pomeron},Q^2) \frac{1}{\sigma_{\rm el}(x)}\int d^2 \vec{b} \left(\left|1-e^{-\frac{1-i \eta}{2} \sigma_{\rm soft}(x) T_A(\vec{b})}\right|^2+e^{-\sigma_{\rm in}(x) T_A(\vec{b})}-e^{-\sigma_{\rm soft}(x) T_A(\vec{b})} \right) \,, 
\label{eq:inc5} 
\end{equation}
where we explicitly indicated the $x$ dependence of the involved cross sections, see the discussion above. Equation~(\ref{eq:inc5}) generalizes Eq.~(\ref{eq:diff5})
by including the nuclear breakup, which also contributes to the total diffraction (rapidity gap) in $\gamma^{\ast}A$ scattering. Similarly to 
Eq.~(\ref{eq:diff6}), one can apply the QCD factorization theorem to Eq.~(\ref{eq:inc5}) and introduce the corresponding diffractive PDFs,
\begin{equation}
\tilde{f}_{i/A}^{D(3)}(x,x_{\Pomeron},Q^2)=f^{D(3)}_{i/p}(x,x_{\Pomeron},Q^2) \frac{1}{\sigma_{\rm el}^i(x)}\int d^2 \vec{b} \left(\left|1-e^{-\frac{1-i \eta}{2} \sigma_{\rm soft}^i(x) T_A(\vec{b})}\right|^2+e^{-\sigma_{\rm in}^i(x) T_A(\vec{b})}-e^{-\sigma_{\rm soft}^i(x) T_A(\vec{b})} \right) \,,
\label{eq:inc6} 
\end{equation}
where $i$ stands for the parton flavor (quark or gluon).

\subsection{Numerical results for nuclear diffractive PDFs}
\label{subsec:results}

\begin{figure}[t]
\centerline{%
\includegraphics[width=9cm]{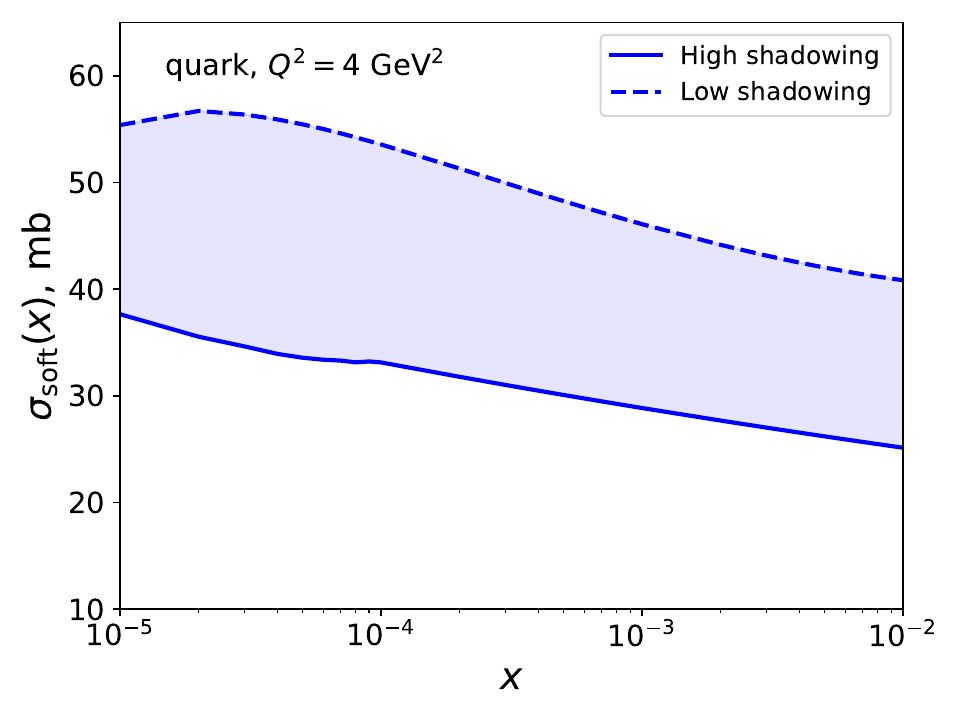}
\includegraphics[width=9cm]{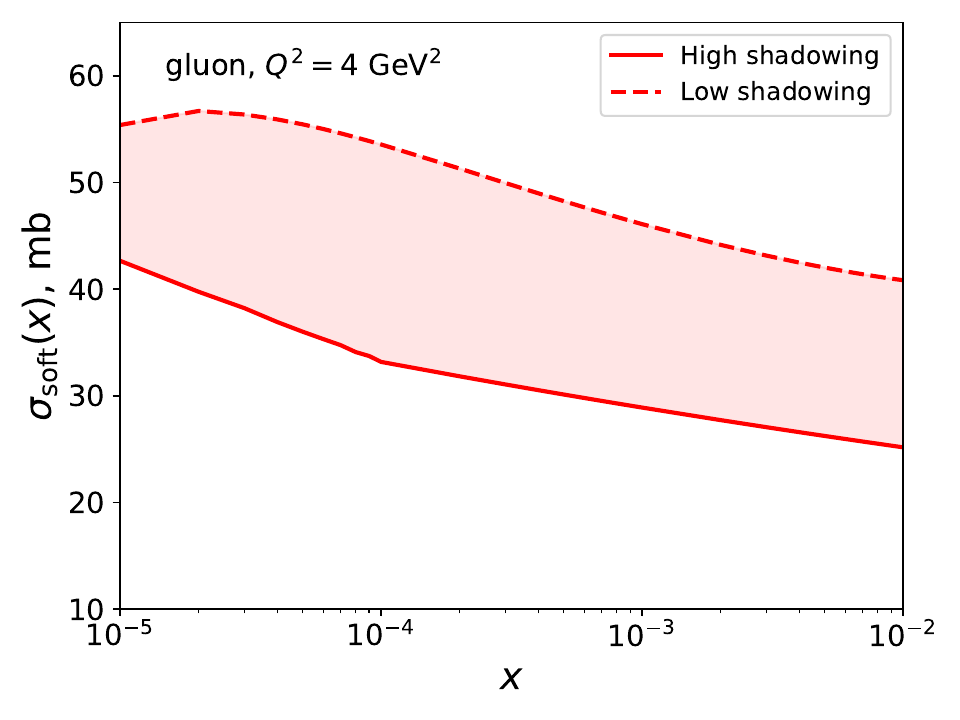}
}
\caption{The band of values for the soft cross section $\sigma_{\rm soft}^i$, which controls the nuclear shadowing effect in the diffractive PDFs
$f_{i/A}^{D(3)}$ and $\tilde{f}_{i/A}^{D(3)}$, as a function $x$ at $Q^2=4$ GeV$^2$. The upper and lower boundaries correspond to the 
``low shadowing'' and ``high shadowing'' scenarios. 
The left and right panels represent the quark and gluon channels, 
respectively.}
\label{fig:sigma}
\end{figure}

As follows from Eqs.~(\ref{eq:diff6}) and (\ref{eq:inc6}), the magnitude of nuclear shadowing in nuclear diffractive PDFs depends on a single parameter -- the cross section $\sigma_{\rm soft}^i$. In this work, we use the results of Ref.~\cite{Frankfurt:2011cs}, where it is modeled using two plausible scenarios for
hadronic fluctuations of the virtual photon. In particular, the lower limit on $\sigma_{\rm soft}^i$ can be estimated using the color dipole model, while
the upper limit on  $\sigma_{\rm soft}^i$ is found by assuming that the relevant hadronic fluctuations are proportional to those of the pion beam.   
The shaded bands in Fig.~\ref{fig:sigma} show the resulting values for $\sigma_{\rm soft}^i$ for quarks (left panel) and gluons (right panel) as a function of $x$ at $Q^2=4$ GeV$^2$. The upper and lower boundaries of this band correspond to the ``low shadowing'' and ``high shadowing'' predictions for 
nuclear PDFs within the leading twist approximation.
Note that $\sigma_{\rm soft}^i$ should be understood as an effective cross section because it involves interference diagrams involving nucleons in
the initial and final nuclear states, see the discussion in Introduction.

\begin{figure}[t]
\centerline{%
\includegraphics[width=9cm]{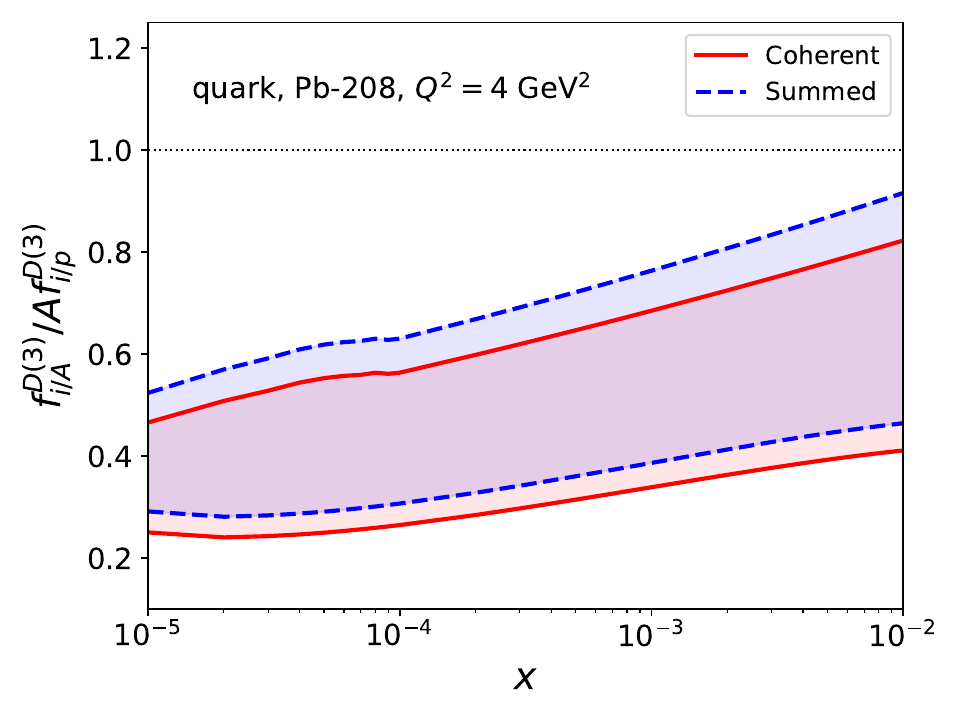}
\includegraphics[width=9cm]{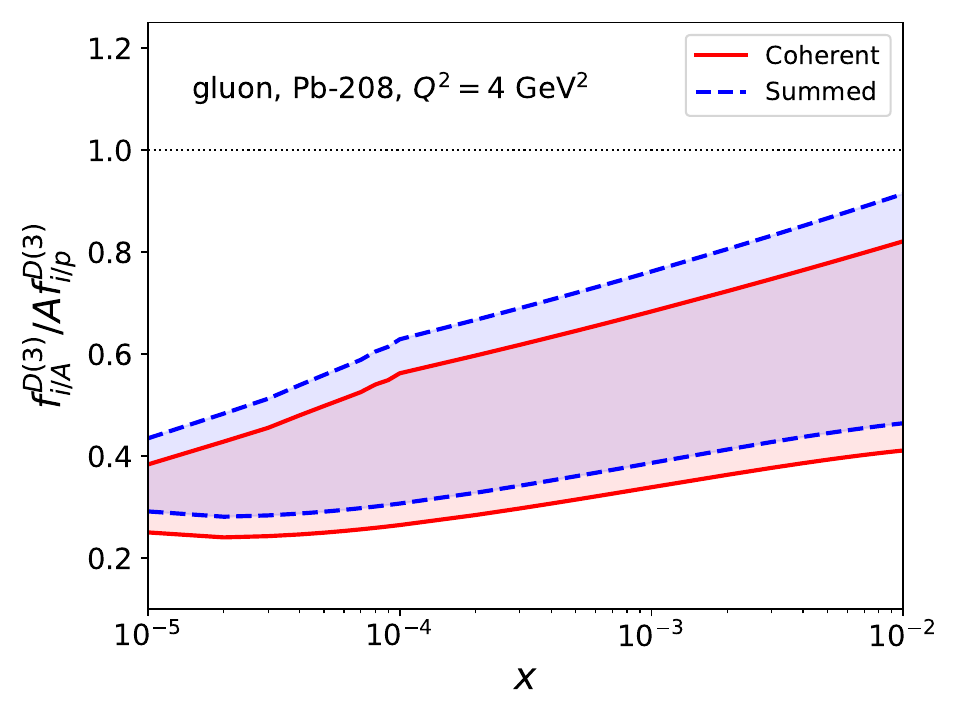}
}
\caption{The LTA predictions for the ratios of the nucleus and proton diffractive PDFs, 
$f_{i/A}^{D(3)}/[Af^{D(3)}_{i/p}]$ and $\tilde{f}_{i/A}^{D(3)}/[Af^{D(3)}_{i/p}]$, as a function of $x$ at $Q^2=4$ GeV$^2$ for $^{208}$Pb.
The left and right panels correspond to the quark and gluon channels, respectively. See text for details.}
\label{fig:Shad_diffraction}
\end{figure}

Figure~\ref{fig:Shad_diffraction} presents the LTA predictions for the ratios of the heavy nucleus ($^{208}$Pb) and proton diffractive PDFs, 
$f_{i/A}^{D(3)}/[Af^{D(3)}_{i/p}]$ and $\tilde{f}_{i/A}^{D(3)}/[Af^{D(3)}_{i/p}]$, as a function of $x$ at $Q^2=4$ GeV$^2$. The ratios are calculated using Eqs.~(\ref{eq:diff6}) (the curves labeled ``Coherent'') and (\ref{eq:inc6}) (the curves labeled ``Summed'')  with $\sigma_{\rm soft}^i$ shown in Fig.~\ref{fig:sigma}. The shaded bands quantify the theoretical uncertainty associated with modeling of $\sigma_{\rm soft}^i$: the upper and lower boundaries correspond to the ``high shadowing'' and ``low shadowing'' predictions, respectively. This order is reversed compared to the LTA predictions for usual nuclear PDFs at small $x$, see the discussion in Sec.~\ref{sec:super-ratio}.
While we separately show the ratios of the quark and gluon diffractive PDFs in the left and right panels, respectively, one can see that 
the dependence on the parton flavor is weak. 

The main feature of the results shown in Fig.~\ref{fig:Shad_diffraction} is a large suppression of the 
presented ratios by the leading twist nuclear shadowing
\begin{equation}
\frac{f_{i/A}^{D(3)}}{Af^{D(3)}_{i/p}} \approx \frac{\tilde{f}_{i/A}^{D(3)}}{Af^{D(3)}_{i/p}} \approx \frac{F^{D(3)}_{2A}}{AF^{D(3)}_{2p}}  \approx \frac{\tilde{F}^{D(3)}_{2A}}{AF^{D(3)}_{2p}} \approx 0.5
\label{eq:ratio_diff}
\end{equation}
at $x=10^{-3}$, $Q^2=4$ GeV$^2$ and independently of $x_{\Pomeron}$ provided that it is small. To appreciate its magnitude, one should compare these results
with the impulse approximation (IA), which is obtained by expanding Eqs.~(\ref{eq:diff6}) and (\ref{eq:inc6}) in powers of $\sigma_{\rm soft}^i$ and keeping 
the leading terms,
\begin{eqnarray}
\frac{f_{i/A}^{D(3)}}{Af^{D(3)}_{i/p}}  & \approx &  \frac{F^{D(3)}_{2A}}{AF^{D(3)}_{2p}} = \frac{4 \pi B_{\rm diff}}{A} \int d^2 \vec{b}\, (T_A(\vec{b}))^2 
=\frac{B_{\rm diff}}{A} \int dt F_A^2(t)= 4.3 \,, \nonumber\\
\frac{\tilde{f}_{i/A}^{D(3)}}{Af^{D(3)}_{i/p}}  & \approx &  \frac{\tilde{F}^{D(3)}_{2A}}{AF^{D(3)}_{2p}} = \frac{4 \pi B_{\rm diff}}{A} \int d^2 \vec{b}\, (T_A(\vec{b}))^2 +1= \frac{B_{\rm diff}}{A} \int dt F_A^2(t)+1=5.3 \,,
\label{eq:ratio_diff2}
\end{eqnarray}
where $F_A(t)$ is the nuclear form factor. This numerical estimate is obtained using the realistic nuclear density $\rho_A$ for $^{208}$Pb~\cite{DeVries:1987atn}. 
Since the diffractive slope $B_{\rm diff}$ has been extracted from the HERA data with ${\cal O}(15\%)$ 
uncertainties~\cite{H1:2006uea}, a similar uncertainty should be assigned to the values on the right-hand side of 
Eq.~(\ref{eq:ratio_diff2}).
Note also that in our analysis, we systematically neglected a possible, small contribution of the Pomeron spin-flip amplitude~\cite{Alberi:1981af} at small $|t| < 3/R_A^2$, which in principle 
somewhat decreases the value of $B_{\rm diff}$.

\section{Probability of diffraction in DIS on nuclei in leading twist approximation}
\label{sec:super-ratio}

As we discussed in Introduction, the ratio of the diffractive and total cross sections for heavy nuclei and the proton in lepton-nucleus DIS is often positioned as an observable 
sensitive to the QCD dynamics at small $x$ and, in particular, to the phenomenon of saturation.
The leading twist approximation also makes predictions for this ratio at the level of structure functions and quark and gluon PDFs.

We start with a brief recapitulation of the LTA predictions for usual nuclear PDFs.
In the small-$x$ limit, LTA allows one to express the ratio of heavy nucleus and free nucleon PDFs in the following compact form~\cite{Frankfurt:2011cs},
\begin{equation}
\frac{f_{i/A}(x,Q^2)}{A f_{i/p}(x,Q^2)}=\lambda^i(x)+(1-\lambda^i(x)) \frac{2}{A \sigma_{\rm soft}^i(x)} \Re e \int d^2 \vec{b} \left(1-e^{-\frac{1-i\eta}{2} \sigma_{\rm soft}^i(x) T_A(\vec{b})}\right) \,,
\label{eq:Rg}
\end{equation}
where 
\begin{equation}
\lambda^i(x)=1-\frac{\sigma_2^i(x)}{\sigma^i_{\rm soft}(x)} \,,
\label{eq:lambda}
\end{equation}
and 
\begin{equation}
\sigma_2^i(x)=\frac{16 \pi B_{\rm diff}}{(1+\eta^2) xf_{i/p}(x,Q_0^2)} \int_x^{0.1} dx_{\Pomeron} \beta f_{i/p}^{D(3)}(x,\beta=x/x_{\Pomeron},Q_0^2)  \,.
\label{eq:sigma2}
\end{equation}
In Eq.~(\ref{eq:sigma2}), the usual and diffractive PDFs are evaluated at $Q_0^2=4$ GeV$^2$.
Equation~(\ref{eq:Rg}) and the parameter $\lambda^i(x)$ have the following transparent physical interpretation in the space-time
picture of high energy hadron-nucleus scattering.  Hadronic fluctuations of the virtual photon can be modeled 
as a superposition of two states: the point-like fluctuation, which interacts with nucleons with a vanishingly small cross section and whose probability is $\lambda^i(x) \leq 1$, and the state interacting with target nucleons with the effective cross section $\sigma_{\rm soft}^i$, whose probability is $1-\lambda^i(x)$~\cite{Frankfurt:1998ym}. 
It leads to a two-component expression for the 
effect of nuclear shadowing given by a sum of the point-like term, which has no nuclear attenuation, and the term proportional to the ratio of the total
nuclear and nucleon cross sections. Expanding Eq.~(\ref{eq:Rg}) in powers of $\sigma_{\rm soft}^i$, one finds that the interaction with two nucleons (the dominant contribution in the weak nuclear density limit) is driven by the cross section $\sigma_2^i$, while the strength of the interaction with
$N \geq 3$ nucleons is determined by $\sigma_{\rm soft}^i$.

\begin{figure}[t]
\centerline{%
\includegraphics[width=9cm]{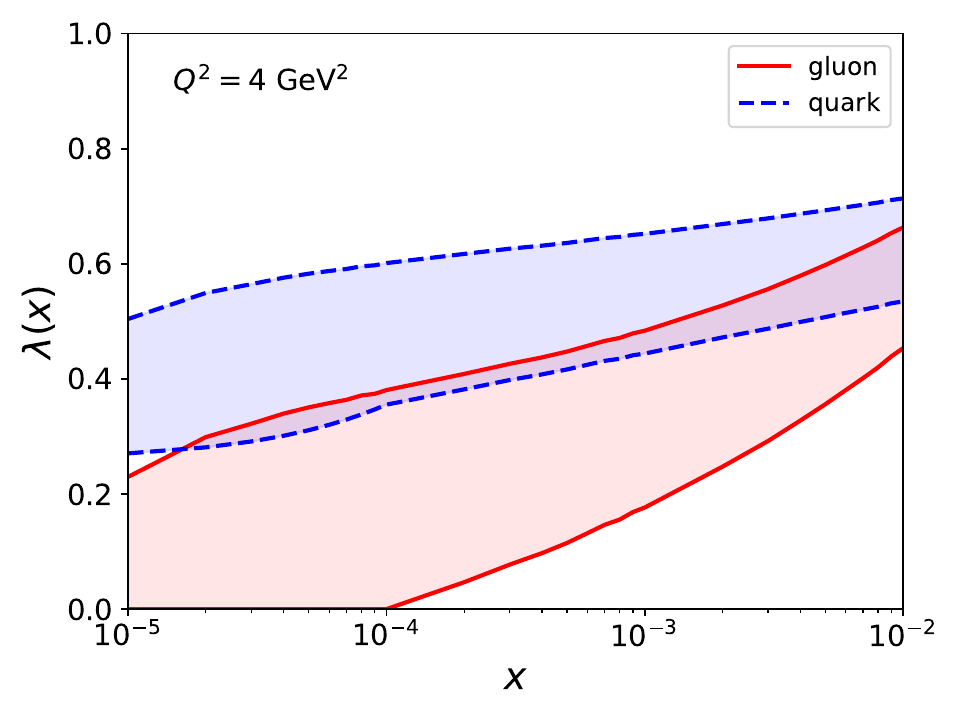}
\includegraphics[width=9cm]{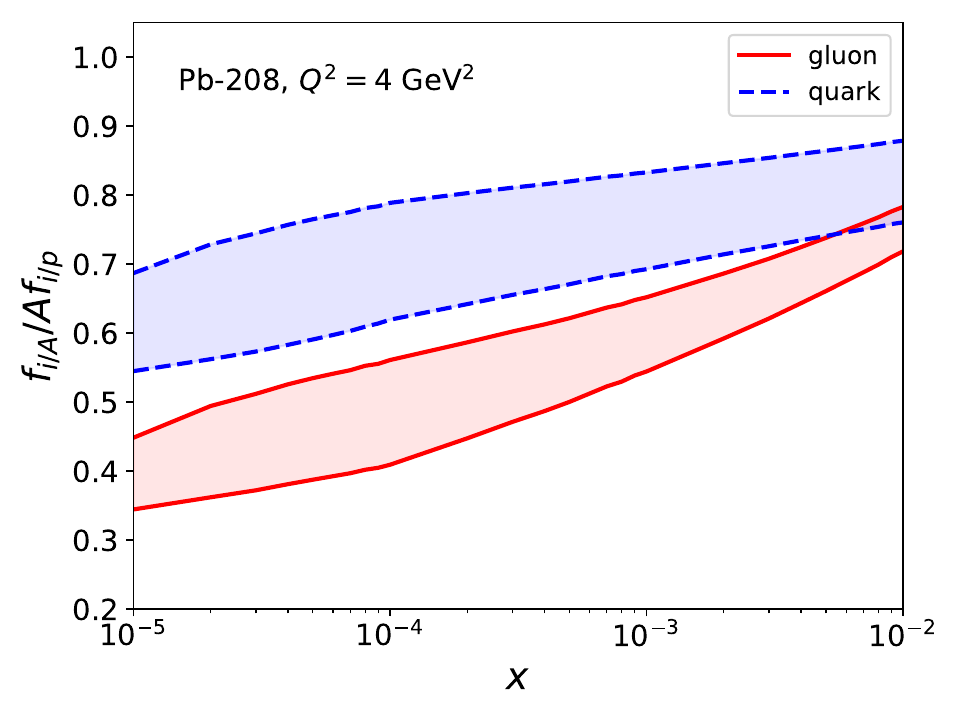}
}
\caption{(Left) The probability of point-like configurations $\lambda^i(x)$ for quarks and gluons as a function of $x$ at $Q^2=4$ GeV$^2$.
(Right) 
The ratios of the nucleus and proton PDFs $f_{i/A}/(Af_{i/p})$ for quarks and gluons as a function of $x$ at $Q^2=4$ GeV$^2$ for $^{208}$Pb.
In both panels, the uncertainty bands originate from those for the soft cross section $\sigma_{\rm soft}^i$, see Fig.~\ref{fig:sigma}.
The upper and lower boundaries correspond to the ``low shadowing' and ``high shadowing'' scenarios, respectively.
}
\label{fig:Rg}
\end{figure}

The left panel of Fig.~\ref{fig:Rg} shows the probability of point-like configurations $\lambda^i(x)$ for quarks (dashed lines) and gluons (solid lines) as a function of $x$ at $Q^2=4$ GeV$^2$. The spread of predictions is characterized by the shaded bands, which originate from the theoretical uncertainty in the value of $\sigma_{\rm soft}^i(x)$, see Fig.~\ref{fig:sigma}. 
As expected, $\lambda^i$ decreases with a decrease of $x$ because the cross sections of hadronic fluctuations (dipole cross sections) increase with 
a decrease of $x$, which reduces the probability of point-like configurations.
One can see from this figure 
that $\lambda^{\rm gluon} < \lambda^{\rm quark}$ because in the space-time picture used above, it is generally expected that 
hadronic fluctuations associated with the gluon nuclear shadowing have on average larger cross sections than those responsible for the 
quark nuclear shadowing. For $x < 10^{-4}$, we assume that $\lambda^{\rm gluon}=0$ because one closely approaches the black limit, see Eq.~(\ref{eq:sigma_bdl}) and its discussion below.
The relatively large values of $\lambda^i$ for quarks, e.g., compare to $\lambda=0.2$ in Ref.~\cite{Frankfurt:1998ym}, in an artifact
of the two-component model for nuclear shadowing. A more detailed modeling of the interaction with $N \geq 3$ nucleons in Eq.~(\ref{eq:Rg}), for instance, in terms of two effective cross sections, is expected to lower the values of $\lambda^i$.

The right panel of Fig.~\ref{fig:Rg} presents the LTA results for the ratios of nuclear ($^{208}$Pb) and proton PDFs $f_{i/A}/(A f_{i/p})$
as a function of $x$ at $Q^2=4$ GeV$^2$. The red solid lines correspond to the gluon PDFs, and the blue dashed lines are for the quark PDFs. 
As before, the uncertainty bands quantify the theoretical uncertainty of the LTA approach associated with modeling of $\sigma_{\rm soft}^i$. 
One can see from this figure that in the small-$x$ limit, the gluon nuclear shadowing is significantly larger than the quark shadowing.
It is a direct consequence of the connection between nuclear shadowing in lepton-nucleus DIS and diffraction in lepton-proton DIS
and the phenomenological result that the gluon diffractive PDF of the proton is much larger than those of quarks (in other words, 
the perturbative Pomeron is made mostly of gluons). Note that the LHC data on coherent $J/\psi$ photoproduction in Pb-Pb UPCs at 
$\sqrt{s_{NN}}=5.02$ TeV~\cite{CMS:2023snh,ALICE:2023jgu} fall inside the error band for $g_{A}(x)/[Ag_{p}(x)]$ for $x < 10^{-4}$.

Note that the LTA error bands in Fig.~\ref{fig:Rg} (right panel) present their conservative, but realistic estimate.
Other sources of uncertainties include $15$\% experimental errors in the value of the slope parameter $B_{\rm diff}$~\cite{H1:2006uea}
and uncertainties of diffractive PDFs $f_{i/p}^{D(3)}$. While the 2006 H1 diffractive PDFs that we use for our numerical calculations do not provide them~\cite{H1:2006zyl}, they have been estimated in more recent QCD analyses of diffractive PDFs of the proton, see, e.g.~\cite{Salajegheh:2022vyv,Goharipour:2018yov,Salajegheh:2023jgi}.
We have checked numerically that the use of the SKMHS23 diffractive PDFs with 16 error PDFs~\cite{Salajegheh:2023jgi} leads to approximately $15$\% uncertainties
in the calculated values of $\sigma_2^i(x)$ for quarks and 20\% for gluons. However, since $\sigma_2^i(x)$ and $\sigma_{\rm soft}^i$ are correlated, 
propagation of $\sigma_2^i(x)$ uncertainties in Eq.~(\ref{eq:Rg}) results in the uncertainties for $f_{i/A}(x,Q^2)/[A f_{i/p}(x,Q^2)]$, which are smaller than those due to modeling of $\sigma_{\rm soft}^i$ shown by the shaded bands in Fig.~\ref{fig:Rg}.

The magnitude of the small-$x$ gluon nuclear shadowing in the ``high shadowing'' case in Fig.~\ref{fig:Rg} approaches the limiting 
value allowed by $S$-channel unitarity or the black disk limit (BDL) for the proton. In this limit, the cross section fluctuations vanish leading to $\lambda^i(x)=0$, 
and all fluctuations interact with the maximal cross section
\begin{equation}
\sigma_{\rm soft}^i(x)=\sigma_{\rm 2}^i(x)= \sigma_{\rm max}=8 \pi B_{\rm diff}\approx 60 \ {\rm mb} \,.
\label{eq:sigma_bdl}
\end{equation}
Here we used Eq.~(\ref{eq:sigma2}) and the fact that in BDL, the diffractive cross section is half of the total cross section, which at the level of PDFs means that
$\int dx_{\Pomeron} \beta f_{i/p}^{D(3)}=(1/2)xf_{i/p}$. Substituting these values in Eq.~(\ref{eq:Rg}), one obtains for $^{208}$Pb
\begin{equation}
\frac{g_A(x,Q^2)}{A g_{p}(x,Q^2)}_{|{\rm BDL}} =\frac{2}{A \sigma_{\rm max}} \int d^2 \vec{b} \left(1-e^{-\frac{1}{2} \sigma_{\rm max} T_A(\vec{b})}\right)= 0.27 \,.
\label{eq:Rg_max}
\end{equation}
In this estimate we also used $\eta=0$ because the interaction is purely absorptive and the scattering amplitude is purely imaginary in BDL. Note that
the calculation of nuclear shadowing with $\lambda^i(x)=0$ corresponds to the eikonal approximation giving the largest nuclear suppression.

The ratios of the diffractive and usual structure functions and PDFs represent the probability of diffraction in DIS (for a given partonic channel).
Combining Eq.~(\ref{eq:Rg}) with Eqs.~(\ref{eq:diff6}) and (\ref{eq:inc6}), one readily finds the LTA predictions for ratios of the diffractive and usual PDFs in a heavy nucleus and the proton,
\begin{equation}
R_{A/p}^{\rm coh}=\frac{f_{i/A}^{D(3)}(x,x_{\Pomeron},Q^2)/f_{i/A}(x,Q^2)}{f_{i/p}^{D(3)}(x,x_{\Pomeron},Q^2)/f_{i/p}(x,Q^2)}=\frac{\sigma_{\rm soft}^i(x)}{\sigma^i_{\rm el}(x)} \frac{\int d^2 \vec{b} \left|1-e^{-\frac{1-i\eta}{2} \sigma_{\rm soft}^i(x) T_A(\vec{b})} \right|^2}
{2(1-\lambda^i(x)) \Re e \int d^2 \vec{b} \left(1-e^{- \frac{1-i\eta}{2} \sigma_{\rm soft}^i(x)T_A(\vec{b})}\right)+\lambda^i(x) A \sigma_{\rm soft}^i(x)}
\label{eq:P_Ap}
\end{equation}
and
\begin{equation}
R_{A/p}=\frac{\tilde{f}_{i/A}^{D(3)}(x,x_{\Pomeron},Q^2)/f_{i/A}(x,Q^2)}{f_{i/p}^{D(3)}(x,x_{\Pomeron},Q^2)/f_{i/p}(x,Q^2)}=\frac{\sigma_{\rm soft}^i(x)}{\sigma^i_{\rm el}(x)} \frac{\int d^2 \vec{b} \left(\left|1-e^{-\frac{1-i\eta}{2} \sigma_{\rm soft}^i(x) T_A(\vec{b})} \right|^2
+e^{-\sigma_{\rm in}^i(x) T_A(\vec{b})}-e^{-\sigma_{\rm soft}^i(x) T_A(\vec{b})}\right)}
{2(1-\lambda^i(x)) \Re e \int d^2 \vec{b} \left(1-e^{- \frac{1-i\eta}{2} \sigma_{\rm soft}^i(x)T_A(\vec{b})}\right)+\lambda^i(x) A \sigma_{\rm soft}^i(x)} \,.
\label{eq:P_Ap_tot}
\end{equation}
Note that in LTA, these ratios do not depend on the Pomeron momentum fraction $x_{\Pomeron}$ for small values of $x_{\Pomeron}$.

\begin{figure}[t]
\centerline{%
\includegraphics[width=9cm]{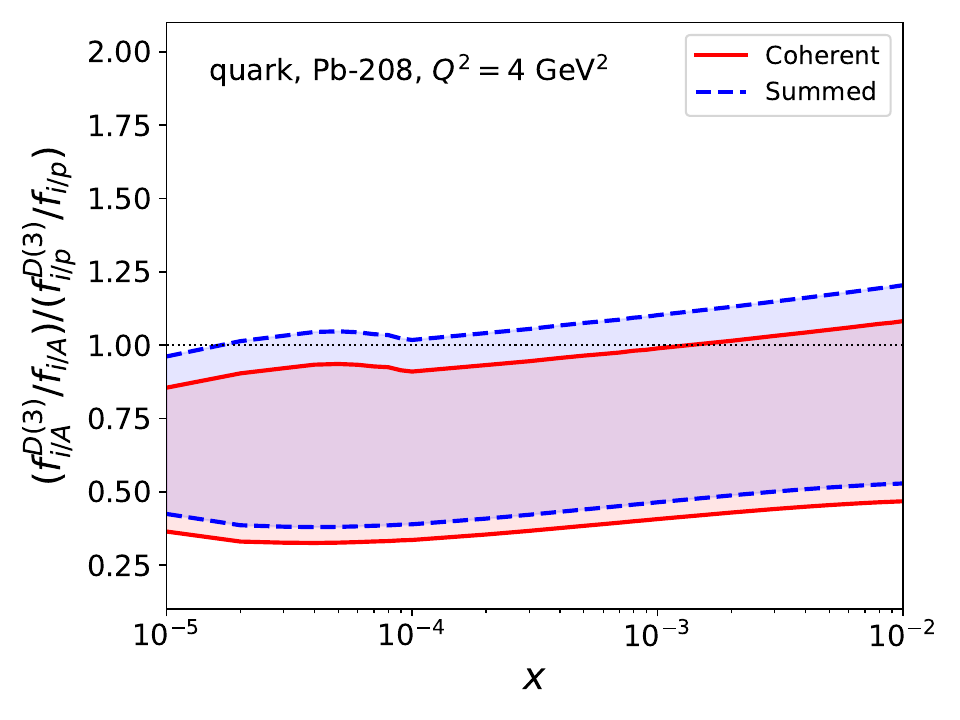}
\includegraphics[width=9cm]{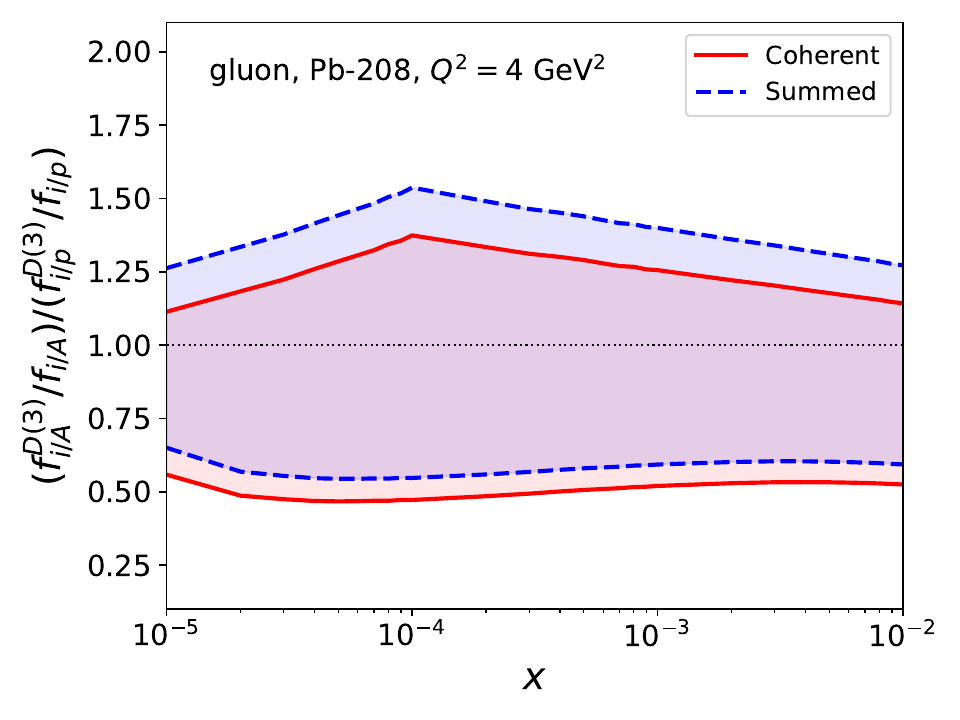}
}
\caption{The ratios of the diffractive and usual PDFs for $^{208}$Pb and the proton, 
$(f_{i/A}^{D(3)}/f_{i/A})/(f_{i/p}^{D(3)}/f_{i/p})$ and $(\tilde{f}_{i/A}^{D(3)}/f_{i/A})/(f_{i/p}^{D(3)}/f_{i/p})$, see Eqs.~(\ref{eq:P_Ap}) and (\ref{eq:P_Ap_tot}), as a function of $x$ at $Q^2=4$ GeV$^2$ for $^{208}$Pb.
The left and right panels are for the quark and gluon channels, respectively. The upper and lower curves correspond to the ``high shadowing'' and ``low shadowing'' scenarios. 
The shaded bands represent theoretical uncertainties of the 
LTA predictions. See text for details.}
\label{fig:Shad_diffraction_PAp}
\end{figure}

Figure~\ref{fig:Shad_diffraction_PAp} shows the LTA predictions for the ratios $R_{A/p}^{\rm coh}$ and $R_{A/p}$ as functions of $x$ at $Q^2=4$ GeV$^2$ for $^{208}$Pb.
These results can be understood by combining those shown in Figs.~\ref{fig:Shad_diffraction} 
and \ref{fig:Rg}. For the quark PDFs (left panel), the ratios $R_{A/p}^{\rm coh} \approx R_{A/p} \approx 0.5$ for the ``low shadowing'' scenario
and $R_{A/p}^{\rm coh} \approx R_{A/p} \approx 1$ in the ``high shadowing'' case. This behavior is largely driven by the usual nuclear PDFs,
whose relative suppression is weaker (``low shadowing'' due to the large probability of point-like configurations $\lambda^i$) or similar (``strong shadowing'' due to smaller $\lambda^i$) compared to that of the nuclear diffractive PDFs, see Fig.~\ref{fig:Rg}.

In the case of gluon PDFs (right panel), the ratios $R_{A/p}^{\rm coh} \approx R_{A/p} \approx 0.5$ in the ``low shadowing'' case and $R_{A/p}^{\rm coh} \approx R_{A/p} \approx 1.2-1.3$ for ``high shadowing''. Similarly to the quark case, it is mostly controlled by the amount of nuclear shadowing in the usual gluon distribution, which is determined by the soft cross section $\sigma_{\rm soft}^i$ and the fraction of point-like configurations $\lambda^i$. 

It is important to note that the ``high shadowing'' curves in Fig.~\ref{fig:Shad_diffraction_PAp} lie above the ``low shadowing'' predictions, 
which is opposite to the trend of the $f_{i/A}/(A f_{i/p})$ ratios. It can be understood by examining the structure of the expressions in Eqs.~(\ref{eq:diff6}),
(\ref{eq:inc6}) and (\ref{eq:Rg}), which show that smaller values of $\sigma_{\rm soft}^i$ corresponds to larger $f_{i/A}^{D(3)}/f_{i/p}^{D(3)}$ 
and $\tilde{f}_{i/A}^{D(3)}/f_{i/p}^{D(3)}$ and smaller $f_{i/A}/(A f_{i/p})$ (for $\lambda^i \neq 0$).
This also explains the very large spread of the LTA predictions for $R_{A/p}$ and $\tilde{R}_{A/p}$ because variations of $\sigma_{\rm soft}^i$ affect
$f_{i/A}^{D(3)}$ and $f_{i/A}$ in an opposite way.

\begin{figure}[t]
\centerline{%
\includegraphics[width=10cm]{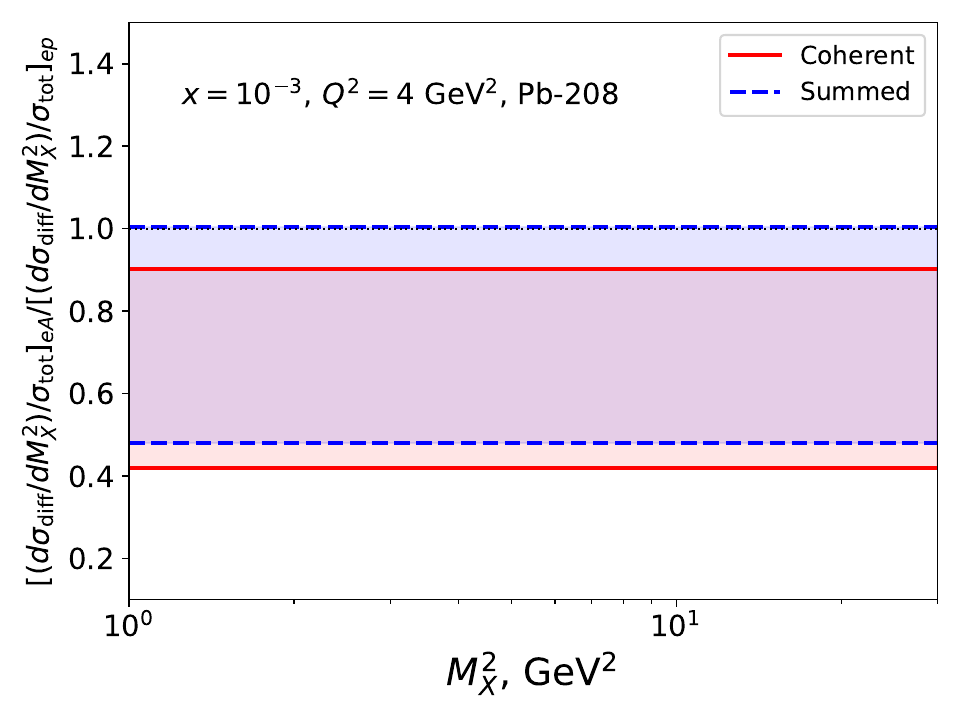}}
\caption{The LTA predictions for the ratio of the diffractive and total cross sections for a heavy nucleus and the proton, $[(d\sigma_{\rm diff}/dM_X^2)/\sigma_{\rm tot}]_{eA}/[(d\sigma_{\rm diff}/dM_X^2)/\sigma_{\rm tot}]_{ep}$,  as a function of $M_X^2$ at $x=10^{-3}$ and $Q^2=4$ GeV$^2$ for $^{208}$Pb.
The red solid and blue dashed curves correspond to the coherent and summed (coherent plus quasi-elastic) nuclear scattering, respectively.
The shaded bands represent theoretical uncertainties of the 
LTA predictions.
}
\label{fig:Shad_diffraction_Mx}
\end{figure}

Equipped with these results, we can now turn to the ratio of the diffractive and total cross sections for a heavy nucleus and the proton, $R_{A/p}=[(d\sigma_{\rm diff}/dM_X^2)/\sigma_{\rm tot}]_{eA}/[(d\sigma_{\rm diff}/dM_X^2)/\sigma_{\rm tot}]_{ep}$. As we discussed in Introduction,
this double ratio is positioned in the literature~\cite{Kovchegov:2023bvy,Accardi:2012qut} as a sensitive observable to distinguish between the leading twist and saturation approaches. Figure~\ref{fig:Shad_diffraction_Mx} shows the LTA predictions for $[(d\sigma_{\rm diff}/dM_X^2)/\sigma_{\rm tot}]_{eA}/[(d\sigma_{\rm diff}/dM_X^2)/\sigma_{\rm tot}]_{ep}$ as a function of $M_X^2$ at $x=10^{-3}$ and $Q^2=4$ GeV$^2$ for $^{208}$Pb. 
The red solid and blue dashed curves correspond to the coherent and summed (coherent plus quasi-elastic) nuclear scattering, respectively, and
the shaded bands represent the LTA theoretical uncertainties.
These predictions are obtained by combining the results of Figs.~\ref{fig:Shad_diffraction} and \ref{fig:Rg}
with the next-to-leading order (NLO) perturbative QCD expressions for the reduced inclusive and diffractive cross sections (structure functions); they are consistent with the LTA
results in~\cite{Kovchegov:2023bvy,Accardi:2012qut}. Note that the curves for $R_{A/p}$ lie slightly below those for light quarks
in the left panel of Fig.~\ref{fig:Shad_diffraction_PAp} and noticeably lower than those for gluons in the right panel of Fig.~\ref{fig:Shad_diffraction_PAp}: this is the effect of both the NLO coefficient functions, where the gluon contribution enters as a correction 
suppressed by the strong coupling constant $\alpha_s(Q^2)$ and which effectively probe somewhat larger values of $x$, and the small contribution of valence and charm quarks to $\sigma_{\rm tot}$ increasing the denominator of $R_{A/p}$.

The main feature of the LTA 
predictions in Fig.~\ref{fig:Shad_diffraction_Mx} is that $R_{A/p} < 1$ both in the coherent and summed cases because of the large leading twist nuclear shadowing strongly suppressing the diffractive cross section. It should be contrasted with $R_{A/p} \approx 1.5-2$ predicted in the gluon saturation
framework~\cite{Kowalski:2007rw,Lappi:2023frf}, where the enhancement of $R_{A/p}$ above unity is driven by the nuclear enhancement of the saturation scale. Note that the LTA predictions are flat in $M_X^2$ for not too large $M_X$ because the nuclear suppression, which is driven by 
the cross section $\sigma_{\rm soft}^i(x)$ in Eqs.~(\ref{eq:diff6}) and (\ref{eq:inc6}), is assumed to be independent on $x_{\Pomeron}$ and $M_X^2$.

\begin{figure}[t]
\centerline{%
\includegraphics[width=9cm]{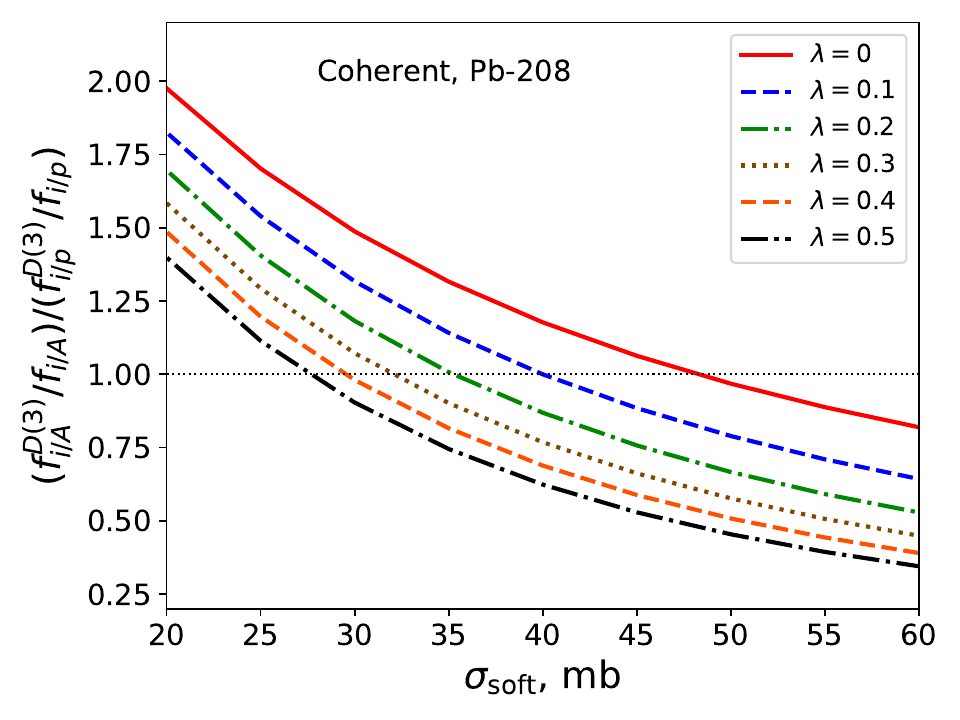}
\includegraphics[width=9cm]{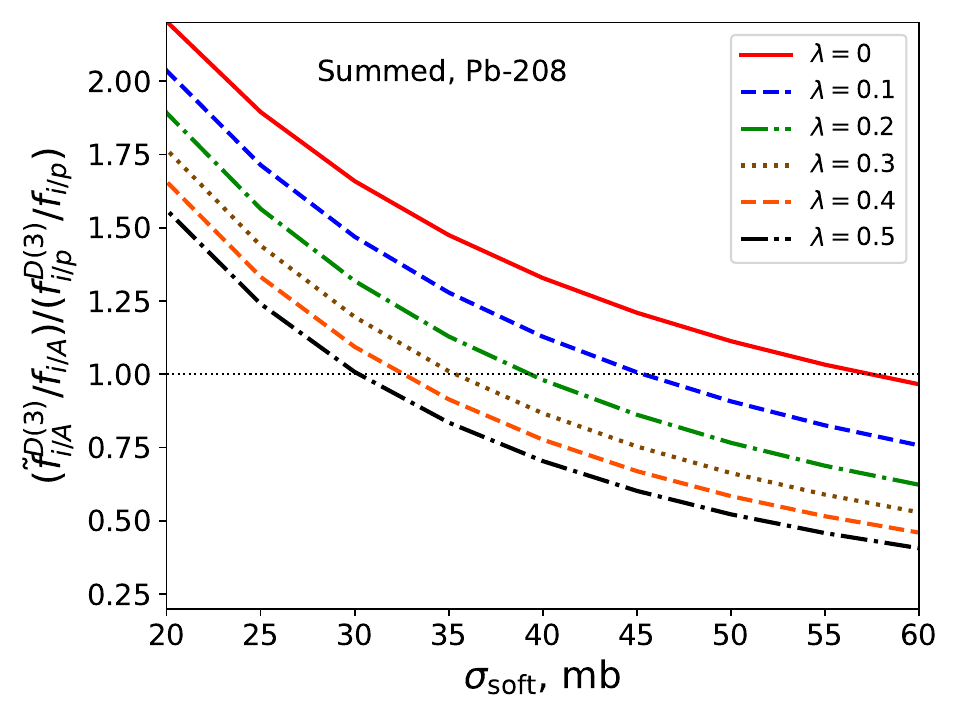}}
\caption{The ratios of the diffractive and usual PDFs (structure functions) for $^{208}$Pb and the proton, $R_{A/p}^{\rm coh}$ and $R_{A/p}$,
see Eqs.~(\ref{eq:P_Ap}) and (\ref{eq:P_Ap_tot}),
as a function of $\sigma_{\rm soft}^i$ for different $\lambda^i$. The left and right panels correspond to the purely coherent and
summed (elastic plus quasi-elastic) nuclear scattering, respectively.}
\label{fig:Shad_diffraction_PAp_sigma3}
\end{figure}

To further illustrate this discussion, in Fig.~\ref{fig:Shad_diffraction_PAp_sigma3} we show $R_{A/p}^{\rm coh}$ and $R_{A/p}$ 
as functions of $\sigma_{\rm soft}^i$ for different choices of $\lambda^i$. One can see from the figure that in the considered interval of $\sigma_{\rm soft}^i$, whether $R_{A/p}^{\rm coh}$ and  $R_{A/p}$ are suppressed or enhanced above unity depends strongly on the values of $\lambda^i$. For example, for 
$\sigma_{\rm soft}^i=40$ mb, $0.7 \leq R_{A/p} \leq 1.3$ for $0.5 \geq \lambda^i \geq 0$.

It is instructive to compare the LTA predictions shown in Figs.~\ref{fig:Shad_diffraction_PAp} and \ref{fig:Shad_diffraction_PAp_sigma3} with those of BDL for the proton, see Eq.~(\ref{eq:sigma_bdl}).
In this limit, $\sigma_{\rm el}^i(x)=\sigma_{\rm in}^i(x)=(1/2) \sigma_{\rm soft}^i(x)=(1/2) \sigma_{\rm max}$ and one obtains
\begin{equation}
\frac{f_{i/A}^{D(3)}(x,x_{\Pomeron},Q^2)/f_{i/A}(x,Q^2)}{f_{i/p}^{D(3)}(x,x_{\Pomeron},Q^2)/f_{i/p}(x,Q^2)}_{|\rm BDL}=\frac{\int d^2 \vec{b} \left(1-e^{-\frac{1}{2} \sigma_{\rm max} T_A(\vec{b})}\right)^2}{\int d^2 \vec{b} \left(1-e^{-\frac{1}{2} \sigma_{\rm max} T_A(\vec{b})}\right)} =0.86 \,, 
\label{eq:PAp_BDL}
\end{equation}
and 
\begin{equation}
\frac{\tilde{f}_{i/A}^{D(3)}(x,x_{\Pomeron},Q^2)/f_{i/A}(x,Q^2)}{f_{i/p}^{D(3)}(x,x_{\Pomeron},Q^2)/f_{i/p}(x,Q^2)}_{|\rm BDL}=\frac{\int d^2 \vec{b} \left[\left(1-e^{-\frac{1}{2} \sigma_{\rm max} T_A(\vec{b})}\right)^2
+e^{-\frac{1}{2} \sigma_{\rm max} T_A(\vec{b})}-e^{-\sigma_{\rm max} T_A(\vec{b})}
 \right]}{\int d^2 \vec{b} \left(1-e^{-\frac{1}{2} \sigma_{\rm max} T_A(\vec{b})}\right)}=1 \,,
\label{eq:PAp_tot_BDL}
\end{equation}
where we used that $\lambda^i(x)=0$ and $\eta=0$ in BDL. 
Note that the limiting values in Eqs.~(\ref{eq:PAp_BDL}) and (\ref{eq:PAp_tot_BDL}) do not depend on parton flavor $i$.
In our estimate in Eq.~(\ref{eq:PAp_BDL}), we employed the realistic nuclear density for $^{208}$Pb~\cite{DeVries:1987atn}. One can see from Fig.~\ref{fig:Shad_diffraction_PAp} that the ``high shadowing'' results for the ratio of gluon PDFs (upper curves, right panel) start to approach  from above the BDL predictions both for $R_{A/p}^{\rm coh}$ and $R_{A/p}$ in the $x \to 10^{-5}$ limit. 
At the same time, when $\sigma_{\rm soft}^i$ has not reached its BDL value and $\lambda^i(x) > 0$, the
LTA predictions for $(f_{i/A}^{D(3)}/f_{i/A})/(f_{i/p}^{D(3)}/f_{i/p})$ and $(\tilde{f}_{i/A}^{D(3)}/f_{i/A})/(f_{i/p}^{D(3)}/f_{i/p})$ in 
Fig.~\ref{fig:Shad_diffraction_PAp} depend on parton flavor $i$ and deviate from flavor-independent predictions of 
Eqs.~(\ref{eq:PAp_BDL}) and (\ref{eq:PAp_tot_BDL}).
In Fig.~\ref{fig:Shad_diffraction_PAp_sigma3}, an approach to BDL is illustrated by the red solid curves in the limit of 
large $\sigma_{\rm soft}^i$.

The LTA predictions in Figs.~\ref{fig:Shad_diffraction_PAp} and \ref{fig:Shad_diffraction_PAp_sigma3} suggest the following picture of energy dependence
for the ratio of the diffractive and total cross sections (total rapidity gap) for a heavy nucleus and the proton. The discussion below refers to
$R_{A/p}$, and similar arguments and estimates are applicable for $R_{A/p}^{\rm coh}$ after a small numerical offset.
\begin{itemize}
\item
Starting at the color transparency (CT) limit of $x \approx 0.1$, where nuclear modifications of nuclear PDFs are expected to be negligibly small, one can use the impulse approximation to find that 
$R_{A/p} \approx 5$, where the enhancement is caused by nuclear coherence. 
\item
As one decreases $x$, the leading twist nuclear shadowing sets in.
Initially, while nuclear shadowing is still shallow (weak), $R_{A/p} \approx 1.5-2$ for $\sigma_{\rm soft}^i=20-30$ mb. 
This corresponds to
the $\rho$ meson-nucleon cross section in the vector meson dominance model and can also serve as an estimate for the quark-antiquark dipole cross section in the color dipole framework. 
It explains the enhancement of $R_{A/p}$ above unity, which does not significantly depend on $\lambda^i$ and parton flavor,
observed in the color dipole framework~\cite{Kowalski:2007rw,Lappi:2023frf,Accardi:2012qut}.

\item
Decreasing $x$, one reaches the full-fledged nuclear shadowing for $\sigma_{\rm soft}^i > 40$ mb.
In this regime, the LTA predictions for $R_{A/p}$ 
depend on modeling of the probability of point-like configurations $\lambda^i$ 
and parton flavor. For gluons, we find that $R_{A/p}$ ranges from 
a mild enhancement $R_{A/p} \approx 1.2-1.3$ to a strong suppression $R_{A/p} \approx 0.5$. In the quark channel and for the ratio
of the diffractive and total cross sections $[(d\sigma_{\rm diff}/dM_X^2)/\sigma_{\rm tot}]_{eA}/[(d\sigma_{\rm diff}/dM_X^2)/\sigma_{\rm tot}]_{ep}$, see Fig.~\ref{fig:Shad_diffraction_Mx}, $R_{A/p}$ is suppressed, 
$R_{A/p} \approx 0.5-1$, which clearly disagrees with predictions of the dipole framework.
One should emphasize that this difference is present in the kinematics of the planned Electron-Ion Collider covering $x \geq 10^{-3}$ for $Q^2 \geq 4$ GeV$^2$~\cite{Accardi:2012qut},
which strengthens the motivation for measurements of $R_{A/p}$.

\item
At the boundary of applicability of the LTA framework, which can be estimated to be reached for $x \lesssim 10^{-5}$ and $\sigma_{\rm soft}^i \approx 60$ mb,
$R_{A/p} \to 1$. This asymptotic LTA prediction contrasts with expectations of the color glass condensate framework, where the nuclear
enhancement of the saturation scale leads to an enhancement of $R_{A/p}$~\cite{Kowalski:2007rw,Lappi:2023frf,Accardi:2012qut}.

\end{itemize}

Note that it is the $R_{A/p}$ ratio, which includes both coherent and quasi-elastic contributions, that approaches unity in BDL rather than 
$R_{A/p}^{\rm coh}$. The difference between $R_{A/p}$ and $R_{A/p}^{\rm coh}$ originates from the contribution of the nucleus edge, 
which is responsible for the BDL prediction of Eq.~(\ref{eq:PAp_BDL}) and which limits the ratio of the nuclear elastic and total cross
sections by the factor of $\approx 0.43$.
For comparison, approximating the nucleus  by a disk of uniform density, $T_A(\vec{b})=\theta (R_A-|\vec{b}|)/(\pi R_A^2)$, where $R_A$ is the effective nucleus radius, one finds that $R_{A/p}^{\rm coh} \to 1$ 
as well as $R_{A/p} = 1$ in the black disk limit.

\section{Leading twist nuclear shadowing and saturation scale}
\label{sec:saturation}
 
The leading twist nuclear shadowing tames the rapid growths of the nuclear gluon distribution at small $x$ and, hence, delays an onset of the non-linear regime
of high parton densities characterized by their 
saturation. One indication of it was presented in the preceding section, where it was shown that the leading twist shadowing significantly suppresses the $R_{A/p}^{\rm coh}$ and $R_{A/p}$ ratios compared to the IA expectations. Another related manifestation of this phenomenon is a significant reduction of the saturation scale by the leading twist nuclear shadowing~\cite{Frankfurt:2022jns}.

The saturation scale $Q_s^2$ can be heuristically defined to be proportional to the gluon density per unit area. Therefore, the ratio of the saturation
scales for a heavy nucleus and the proton can be defined as
\begin{eqnarray}
\frac{Q_{sA}^2(b)}{Q_{sp}^2(b)}&=&\frac{g_A(x,b,Q^2)}{g_p(x,b,Q^2)}=\pi R_p^2 \,\frac{g_A(x,b,Q^2)}{g_p(x,Q^2)} \nonumber\\
&=&\pi R_p^2 \left[\lambda^i(x)T_A(\vec{b}) +(1-\lambda^i(x)) \frac{2}{\sigma_{\rm soft}^i(x)} \Re e \left(1-e^{-\frac{1-i\eta}{2} \sigma_{\rm soft}^i(x) T_A(\vec{b})}\right)\right] \,,
\label{eq:Q_sat}
\end{eqnarray}
where 
$\vec{b}$ is the impact parameter and $g_A(x,b,Q^2)$ and $g_p(x,b,Q^2)$ are the impact parameter dependent nuclear and proton gluon distributions, respectively. For the former, we generalized
Eq.~(\ref{eq:Rg}) to impact parameter dependent nuclear PDFs and used
\begin{equation}
\frac{f_{i/A}(x,b,Q^2)}{f_{i/p}(x,Q^2)}=\lambda^i(x)T_A(\vec{b})+(1-\lambda^i(x)) \frac{2}{\sigma_{\rm soft}^i(x)} \Re e  \left(1-e^{-\frac{1-i\eta}{2} \sigma_{\rm soft}^i(x) T_A(\vec{b})}\right) \,.
\label{eq:Rg_bdep}
\end{equation}
For $g_p(x,b,Q^2)$, we assumed an exponential dependence on the impact parameter,
\begin{equation}
g_p(x,b,Q^2)=\frac{e^{-b^2/R_p^2}}{\pi R_p^2} g_p(x,Q^2) \,,
\label{eq:gp_bdep}
\end{equation}
where $R_p$ is the radius of the gluon density in the proton in the transverse plane. It can be related to the slope of the $t$
dependence of exclusive $J/\psi$ photoproduction on the proton, $B_{J/\psi}$~\cite{Frankfurt:2010ea}
\begin{equation}
R_p^2=2 B_{J/\psi}=9+0.4 \ln (x_0/x) \ {\rm GeV}^{-2} \,,
\label{eq:R_p}
\end{equation}
where in the numerical estimate we used the parametrization of~\cite{Guzey:2013qza} with $x_0=5 \times 10^{-4}$. Note that since $R_p \ll R_A$, 
we used $g_p(x,b=0,Q^2)$ in the second line of Eq.~(\ref{eq:Q_sat}).
One should point out that it is the impact parameter dependent saturation scale $Q_{sA}^2(b)$ rather than the $b$-averaged $Q_{sA}^2$ that quantifies an onset of saturation: while the unitarity bound may be already reached for partial waves (color dipoles) for small $b$,
its effects are washed out in the $b$-averaged case because of the low nuclear density $T_A(\vec{b})$ at the nucleus periphery, 
see, e.g.~\cite{Rogers:2003vi}.

To appreciate the effect of the leading twist nuclear shadowing on $Q_{sA}^2(b)/Q_{sp}^2(b)$, one can compare it with the impulse approximation (IA) estimate, which can be obtained by expanding the right-hand side of Eq.~(\ref{eq:Q_sat}) in powers of $\sigma_{\rm soft}^i$ and keeping the first non-vanishing contribution, 
\begin{equation}
\frac{Q_{sA}^2(b)}{Q_{sp}^2(b)}_{|{\rm IA}}=\pi R_p^2\, T_A(\vec{b}) \,.
\label{eq:Q_sat_IA}
\end{equation}

\begin{figure}[t]
\centerline{%
\includegraphics[width=10cm]{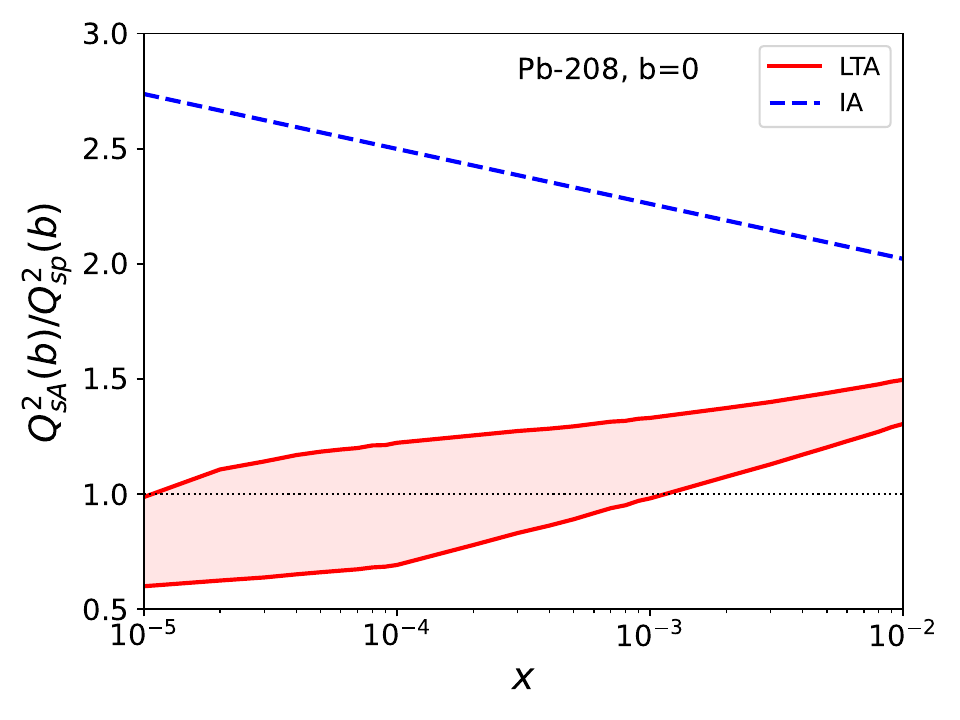}
}
\caption{The LTA predictions for the ratio of the saturation scales in $^{208}$Pb and the proton as a function of $x$ at $b=0$, see Eq.~(\ref{eq:Q_sat}). The blue dashed curve gives the impulse approximation (IA) result of Eq.~(\ref{eq:Q_sat_IA}).
}
\label{fig:Shad_diffraction_bdep}
\end{figure}

Figure~\ref{fig:Shad_diffraction_bdep} shows the LTA predictions for the $Q_{sA}^2(b)/Q_{sp}^2(b)$ ratio of Eq.~(\ref{eq:Q_sat}) as a function of $x$ at $b=0$ for $^{208}$Pb. The shaded band quantifies the theoretical uncertainty associated with the $\sigma_{\rm soft}^i(x)$ cross section, with the upper and lower boundaries corresponding to the ``low shadowing'' and ``high shadowing'' scenarios. For comparison, the IA prediction of Eq.~(\ref{eq:Q_sat_IA}) is given by the blue dashed curve.
One can see from the figure that at small $x$, $Q_{sA}^2(b)/Q_{sp}^2(b)_{|b=0} \approx 1$ (and may even dip below unity at 
$x \to 10^{-5}$) because of the strong $b$-dependent leading twist nuclear shadowing and a modest value of the radius of the
 gluon distribution in the proton
$R_p$ [$R_p \approx 0.6$ fm at $x=10^{-4}$ from Eq.~(\ref{eq:R_p})].
The latter further dilutes the gluon distribution in the transverse plane in a heavy nucleus compared to the proton.
Note that the latter argument is not applicable at $\vec{b} \neq 0$, e.g., $|\vec{b}| \approx 1-2$ fm, and also in the impact-parameter averaged case, where one expects a nuclear enhancement of $Q_{sA}^2/Q_{sp}^2$ tamed by the leading twist nuclear shadowing~\cite{Frankfurt:2022jns}.

As we discussed in Sec.~\ref{sec:super-ratio}, the interaction in the gluon channel in the ``high shadowing'' case  is close to the maximal one corresponding to BDL for very small $x$, see Eqs.~(\ref{eq:sigma_bdl}) and (\ref{eq:Rg_max}).
In this limit, Eq.~(\ref{eq:Q_sat}) at $b=0$ reduces to
\begin{equation}
\frac{Q_{sA}^2(b)}{Q_{sp}^2(b)}_{|b=0} \approx \frac{ 2\pi R_p^2}{\sigma_{\rm soft}^i(x)}  \,.
\label{eq:Q_sat_bdl}
\end{equation}
It shows that $Q_{sA}^2(b)/Q_{sp}^2(b)_{|b=0}$ is determined by the ratio of the cross section associated with the gluon distribution
in the transverse plane in the proton and the soft cross section related to inclusive diffraction on the nucleon. Since the gluon distribution is
known to be rather localized, see Eqs.~(\ref{eq:gp_bdep}) and (\ref{eq:R_p}), $Q_{sA}^2(b)/Q_{sp}^2(b)_{|b=0}$ can dip below unity in the limit, 
where Eq.~(\ref{eq:Q_sat_bdl}) holds.

In this respect, it is important to note that invoking the notion of ``hot gluonic spots'' corresponding to small $R_p$~\cite{Cepila:2017nef,Mantysaari:2016ykx,Mantysaari:2017dwh}  makes 
observation of the gluon saturation in nuclei very challenging.

Therefore, in the picture of high-energy virtual photon-nucleus scattering, where the target nucleus is described in terms of individual 
nucleons and the interaction proceeds via diffractive exchanges with these nucleons, the strong leading twist nuclear shadowing significantly delays an onset of the non-linear regime of saturation in the region, where the leading twist and saturation theoretical descriptions are 
applicable and overlap.
The latter is supported by the observation that $Q_0^2=4$ GeV$^2$ in LTA and $Q_s^2 \sim {\cal O}({\rm few\, GeV}^2)$ for a heavy nucleus 
in the saturation framework.

\section{Conclusions}
\label{sec:conclusions}

Using the leading twist approach to nuclear shadowing, we made detailed predictions for the ratios of diffractive and usual PDFs for a heavy nucleus and the proton, $R^{\rm coh}_{A/p}=(f_{i/A}^{D(3)}/f_{i/A})/(f_{i/p}^{D(3)}/f_{i/p})$ for the purely coherent nuclear DIS and $R_{A/p}=(\tilde{f}_{i/A}^{D(3)}/f_{i/A})/(f_{i/p}^{D(3)}/f_{i/p})$ for the summed (coherent plus quasi-elastic) scattering. We found that $R_{A/p}^{\rm coh} \approx R_{A/p} \approx 0.5-1$ for quarks 
as well as for the ratio of the diffractive and total cross sections $[(d\sigma_{\rm diff}/dM_X^2)/\sigma_{\rm tot}]_{eA}/[(d\sigma_{\rm diff}/dM_X^2)/\sigma_{\rm tot}]_{ep}$
and $R_{A/p}^{\rm coh} \approx R_{A/p} \approx 0.5-1.3$ for gluons 
in a broad range of $x$, $10^{-5} < x < 10^{-2}$. These results are independent of $x_{\Pomeron}$ and 
reaffirm the difference from the results of the dipole model and gluon saturation framework. 
We demonstrated that the magnitude of $R_{A/p}^{\rm coh}$ and $R_{A/p}^{\rm coh}$
is controlled by the cross section of the interaction of hadronic fluctuations of the virtual photon with target nucleons 
$\sigma_{\rm soft}^i$ and its uncertainty, which can also be interpreted in terms of point-like non-interacting fluctuations.
In leads to a natural explanation of different regimes for $R_{A/p}$ (and similarly for $R_{A/p}^{\rm coh}$):
$R_{A/p} \approx 5$ in the color transparency limit of $\sigma_{\rm soft}^i \to 0$;
 $R_{A/p} \approx 1.5-2$ for $\sigma_{\rm soft}^i=20-30$ mb corresponding to typical values used in the dipole model;
 and $R_{A/p} \approx 0.5-1$ (quarks) and $R_{A/p} \approx 0.5-1.3$ (gluons) for $\sigma_{\rm soft}^i=40-50$ mb in the case of the full-fledged leading twist nuclear shadowing for $10^{-5} < x < 10^{-3}$.
 In the black disk limit for the proton, which is estimated to take place at $\sigma_{\rm soft}^i \approx 60$ mb, $R_{A/p}=1$ for the summed cross section
 and $R^{\rm coh}_{A/p}=0.86$ in the case of purely coherent scattering.

Employing an intuitive definition of the saturation scale, we showed that 
the ratio of the saturation scales of a heavy nucleus and proton $Q_{sA}^2(b)/Q_{sp}^2(b) \approx 1$ at small impact parameters $b$. This 
absence of a nuclear enhancement of the saturation scale, which is commonly expected to scale as $A^{1/3}$ based on the  nucleon counting (the so-called ``oomph'' factor~\cite{Accardi:2012qut}), is caused by the strong leading twist 
nuclear shadowing and relative diluteness of the nuclear gluon distribution (nuclear density) in the transverse plane compared to that in the proton.

In general, numerical results presented in this paper indicate that the leading twist nuclear shadowing significantly delays an onset of the non-linear regime of 
gluon saturation in the kinematical region, where the interaction is ``grey'' and where both the leading twist and parton saturation theoretical descriptions are applicable.
Nevertheless, our results strengthen the physics case for measuring $R_{A/p}^{\rm coh}$ and $R_{A/p}$ and their flavor dependence in the EIC kinematics, which
covers $x \geq 10^{-3}$ for $Q^2 \geq 4$ GeV$^2$, 
since it provides an important ingredient for establishing the dynamical mechanism of small-$x$ nuclear shadowing.
Note, however, that to achieve it unambiguously, one needs to study several observables in $\gamma^{\ast} A$ scattering, including charmonium and heavy-flavor jet production, as well as
their counterparts in photon-nucleus scattering in UPCs.

\acknowledgments
The research of V.G.~was funded by the Academy of Finland project 330448, the Center of Excellence in Quark Matter of the Academy of Finland (projects 346325 and 346326), and the European Research Council project ERC-2018-ADG-835105 YoctoLHC. The research of M.S.~was supported by the US Department of Energy Office of Science, Office of Nuclear Physics under Award No. DE- FG02-93ER40771.

\end{document}